%% file: main.tex
\documentclass[acmsmall, screen]{acmart}
\acmConference[PL]{}{2021}{USA}
\acmYear{}

\acmISBN{}
\acmDOI{}
\startPage{1}

\setcopyright{none}

\usepackage{booktabs}
\usepackage{subcaption}

\usepackage{balance}
\usepackage{pifont}
\usepackage{wrapfig}
\usepackage{hyperref}
\usepackage{fancyvrb}
\usepackage{xspace}
\usepackage{algorithmicx}
\usepackage{algpseudocode}
\usepackage{algorithm}
\usepackage{amsfonts}
\usepackage{amsmath}
\usepackage{amsthm}
\usepackage{semantic}
\usepackage{listings}
\usepackage{array}
\usepackage{fancyvrb}
\usepackage{mathpartir}
\usepackage{stmaryrd}
\usepackage{graphicx}
\usepackage{caption}
\usepackage{syntax, etoolbox}
\usepackage{tikz}
\usepackage{proof}
\usepackage{color}
\usepackage{mathtools}
\usepackage{adjustbox}
\usepackage{multirow}
\usepackage{soul} 

\lstdefinestyle{ruler}{
  basicstyle=\footnotesize\sffamily,
}
\lstdefinestyle{rewrite}{
  basicstyle=\scriptsize\sffamily,
  gobble=4,
}
\makeatletter
\lst@AddToHook{Init}{\mathligsoff}
\makeatother

\usepackage[utf8x]{inputenc}
\input{macros}

\begin{document}

\title[Ruler]{Rewrite Rule Inference Using Equality Saturation}

\author{Chandrakana Nandi}
\affiliation{
  \institution{University of Washington}
  \country{USA}
}
\email{cnandi@cs.washington.edu}
\authornote{Both authors contributed equally to this work.}

\author{Max Willsey}
\orcid{0000-0001-8066-4218}
\affiliation{
  \institution{University of Washington}
 \country{USA}
}
\email{mwillsey@cs.washington.edu}
\authornotemark[1]

\author{Amy Zhu}
\affiliation{
  \institution{University of Washington}
  \country{USA}
}
\email{amyzhu@cs.washington.edu }

\author{Yisu Remy Wang}
\affiliation{
  \institution{University of Washington}
  \country{USA}
}
\email{remywang@cs.washington.edu }

\author{Brett Saiki}
\affiliation{
  \institution{University of Washington}
  \country{USA}
}
\email{bsaiki@cs.washington.edu}

\author{Adam Anderson}
\affiliation{
  \institution{University of Washington}
  \country{USA}
}
\email{adamand2@cs.washington.edu}

\author{Adriana Schulz}
\affiliation{
  \institution{University of Washington}
  \country{USA}
}
\email{adriana@cs.washington.edu}

\author{Dan Grossman}
\affiliation{
  \institution{University of Washington}
  \country{USA}
}
\email{djg@cs.washington.edu}

\author{Zachary Tatlock}
\affiliation{
  \institution{University of Washington}
  \country{USA}
}
\email{ztatlock@cs.washington.edu}

 \renewcommand{\shortauthors} {C.Nandi, M. Willsey, A. Zhu, Y. R. Wang, B. Saiki, A. Anderson, A. Schulz, D. Grossman, Z. Tatlock}

\begin{abstract}
  \input{00-abstract}

\end{abstract}


\maketitle

\AtBeginEnvironment{grammar}{\small}
\input{01-introduction}

\input{02-background}

\input{03-algorithm}

\input{04-evaluation}

\input{05-casestudy}

\input{06-ablation}

\input{07-future}
\input{09-related}

\input{10-conclusions}

\begin{acks}
We thank the anonymous reviewers for their thoughtful feedback.
We are grateful to Andrew Reynolds for answering our questions about CVC4's
  rule synthesis component,
  to Pavel Panchekha and Oliver Flatt for helping us run Herbie with \ruler's rules, and
  to Sorawee Porncharoenwase and Jacob Van Geffen for helping with Rosette.
Thanks to Steven S. Lyubomirsky and Bill Zorn for allowing us to run
  experiments on their research servers.
Thanks to Talia Ringer, Martin Kellogg, Ben Kushigian, Gus Henry Smith, and Mike He
  for their feedback on earlier drafts of the paper.
\end{acks}

\bibliography{reference}

\end{document}

%% file: macros.tex
\newcommand{\todo}[1]{{\color{red}[#1]}}

\renewcommand{\max}[1]{{\color{purple}mw: #1}}
\newcommand{\adriana}[1]{{\color{blue}as: #1}}

\newcommand{\amy}[1]{{\color{brown}az: #1}}

\newcommand{\chandra}[1]{{\color{cyan}cn: #1}}

\newcommand{\change}[1]{{#1}}

\definecolor{bp}{rgb}{0.2, 0.2, 0.6}

\newcommand{\tool}{Ruler\xspace}
\newcommand{\ruler}{Ruler\xspace}
\newcommand{\Ruler}{Ruler\xspace}
\newcommand{\egg}{\texorpdfstring{\MakeLowercase{\texttt{egg}}}{\texttt{egg}}\xspace}
\newcommand{\egraph}{\mbox{e-graph}\xspace}
\newcommand{\egraphs}{e-graphs\xspace}
\newcommand{\Egraph}{\mbox{E-graph}\xspace}
\newcommand{\Egraphs}{\mbox{E-graphs}\xspace}

\newcommand{\eclass}{e-class\xspace}
\newcommand{\eclasses}{e-classes\xspace}

\newcommand{\enode}{e-node\xspace}
\newcommand{\enodes}{e-nodes\xspace}

\newcommand{\cvec}{cvec\xspace}

\newcommand{\cvecs}{cvecs\xspace}
\newcommand{\Eqsat}{Equality saturation\xspace}
\newcommand{\eqsat}{equality saturation\xspace}
\newcommand\rewritesto{\ensuremath{\rightsquigarrow}\xspace}
\newcommand\rewritesboth{\ensuremath{\leftrightsquigarrow}\xspace}

\newcommand{\rationals}{{rationals}\xspace}
\newcommand{\bfour}{{bitvector-4}\xspace}
\newcommand{\bthreetwo}{{bitvector-32}\xspace}
\newcommand{\booleans}{{booleans}\xspace}

\newcommand{\mypara}[1]{\paragraph{#1}}

%% file: 00-abstract.tex

Many compilers, synthesizers, and theorem provers rely on
  rewrite rules to simplify expressions or prove equivalences.
Developing rewrite rules can be difficult:
  rules may be subtly incorrect,
  profitable rules are easy to miss, and
  rulesets must be rechecked or extended
  whenever semantics are tweaked.
Large rulesets can also be challenging to apply:
  redundant rules slow down rule-based search
  and frustrate debugging.
  
This paper explores how \eqsat,
  a promising technique that uses \egraphs 
  to \textit{apply} rewrite rules,
  can also be used to \textit{infer} rewrite rules.
\Egraphs
  can compactly represent
  the exponentially large sets of enumerated terms and potential rewrite rules.
We show that \eqsat 
  efficiently shrinks both sets,
  leading to faster synthesis of smaller, more general rulesets.


We prototyped these strategies in a tool dubbed \ruler.
Compared to a similar tool built on CVC4,
  \ruler synthesizes $5.8\times$ smaller 
  rulesets $25\times$ faster
  without compromising on proving power.
In an end-to-end case study,
  we show \ruler-synthesized rules which perform as well
 as those crafted by domain experts, 
 and addressed a longstanding issue in a popular open source tool.

%% file: 01-introduction.tex

\section{Introduction}
\label{sec:intro}

%
%

Rewrite systems transform expressions
  by repeatedly applying a given set of rewrite rules.
Each rule $\ell \rewritesto r$ rewrites
  occurrences of the syntactic pattern $\ell$ to
  instances of another
  \textit{semantically equivalent}
  pattern $r$.
Rewrite systems are effective because they
  combine individually simple rules
  into sophisticated transformations,
  maintain equivalence between rewritten expressions, and
  are easy to extend by adding new rules.


\input{fig-bv.tex}
Many compilers, program synthesizers, and theorem provers
  rely on rewrite systems~\cite{
    haskell, arvind-hw-synth-rw, simplify}.
For example, rewriting is essential for
  improving program analyses and code generation~\cite{
    isel-survey, mlir, halide, tvm}
  and for automating verification~\cite{
    cvc4, z3, isabelle, coq}.
Without rule-based simplification,
  Halide-generated code can
  suffer 26$\times$ slowdown~\cite{julie-halide}
  and
  the Herbie floating-point synthesizer~\cite{herbie}
  can return 10$\times$ larger programs.
 



Several noteworthy projects have developed
  tool-specific techniques for checking or inferring rules~\cite{
    bansal, alive-infer, denali, swapper},
  but 
  implementing a rewrite system
  still generally requires domain experts to
  first manually develop rulesets by trial and error.
Such slow, ad hoc, and error-prone approaches
  hinder design space exploration for new domains
  and discourage updating existing systems.


To address these challenges,
  we propose a simple, domain-general approach
  that uses \eqsat~\cite{eqsat, egg} as
  a rewrite system \textit{on the domain of rewrite rules themselves}
  to quickly synthesize effective rulesets.
  
In the past, tool-specific techniques
  to iteratively infer rewrite rules
  have implicitly adopted a common three-step approach,
  each constructing or maintaining a set:
\begin{enumerate}
    \item Enumerate terms from the given domain to
          build the \textit{term set} $T$.
    \item Select candidate rules from $T \times T$ 
          to build the \textit{candidate set} $C$.
    \item Filter $C$ to select a sound set of
          useful rules to build the \textit{rule set} $R$.
\end{enumerate}
We identify and abstract this workflow
  to provide generic rule inference for user-specified domains.
  
Our key insight is that \emph{what
  makes \eqsat successful in rewrite rule application 
  is also useful for rule inference}.
\Eqsat can simultaneously prove many pairs of terms equivalent
 with respect to a given ruleset.
Our approach uses \eqsat to shrink
  the set $T$ of enumerated terms
  (lowering candidate \textit{generation} cost)
    by merging terms equivalent under $R$, 
    and to shrink the set $C$ of candidate rules
  (lowering candidate \textit{selection} cost)
    by removing rules derivable by $R$.
Thus, it uses the set $R$ of rewrite rules
  to rewrite the next batch of candidate rewrite rules
  \textit{even as $R$ is being synthesized}.

We prototyped these insights in
  a tool dubbed \ruler (\autoref{fig:grammar1}).
Compared to a state-of-the-art 
 rule synthesizer \cite{sat19} built
 into the CVC4 theorem prover~\cite{cvc4},
  Ruler synthesizes smaller rulesets in less time
  without reducing the set of derivable equivalences.
We demonstrate how \ruler
  can generate expert-quality rulesets
  by using it to replace all of
  Herbie's rules for rational numbers,
  uncovering missing rules that
  resolved a known bug in Herbie.


In summary, this paper's contributions include:
\begin{itemize}
  \item A novel rule synthesis algorithm
    that uses \egraphs~\cite{nelson} to
    compactly encode large sets of terms and
    \eqsat to efficiently filter and minimize rulesets
    (\autoref{sec:core}).
    
  \item A generic implementation of
    this algorithm within the 
    \ruler\footnote{
        \ruler will be made open-source and publicly available at
        \texttt{[link redacted for review]}.}
    rewrite rule inference framework
    that synthesizes rules for user-specified domains
    given a grammar and its interpreter.
    
  \item A comparison against a recent CVC4-based rule synthesizer
    that shows \ruler synthesizes $5.8\times$ smaller 
    rulesets $25\times$ faster without compromising
    the deriving power of the rulesets.
    
  \item A case study demonstrating that,
    in an end-to-end application of a real world tool,
    \ruler's automatically generated rulesets
    are as good as manually-crafted expert rules
    (\autoref{sec:end2end}).
    
\end{itemize}

The rest of the paper is organized as follows.
\autoref{sec:back}
 provides background on rewrite systems, 
 \egraphs, and \eqsat.
\autoref{sec:ruler} 
 presents \ruler's core algorithm,
\autoref{sec:eval}
 evaluates our implementation of Ruler against a CVC4-based rule synthesizer.
\autoref{sec:case} presents a case study
 where \ruler-generated rules can competitively replace
 expert-crafted rules in the
 Herbie numerical program synthesizer.
\autoref{sec:ablation} provides several empirical analyses of \tool's algorithm
  and compares different instantiations, verification back-ends, and
  rule discovery strategies.
\autoref{sec:limit} discusses limitations of our approach and 
 opportunities for future work.
\autoref{sec:related} presents related work,
 and \autoref{sec:conclusions} concludes.

%% file: fig-bv.tex
\begin{figure*}[t]
\includegraphics[width=\textwidth]{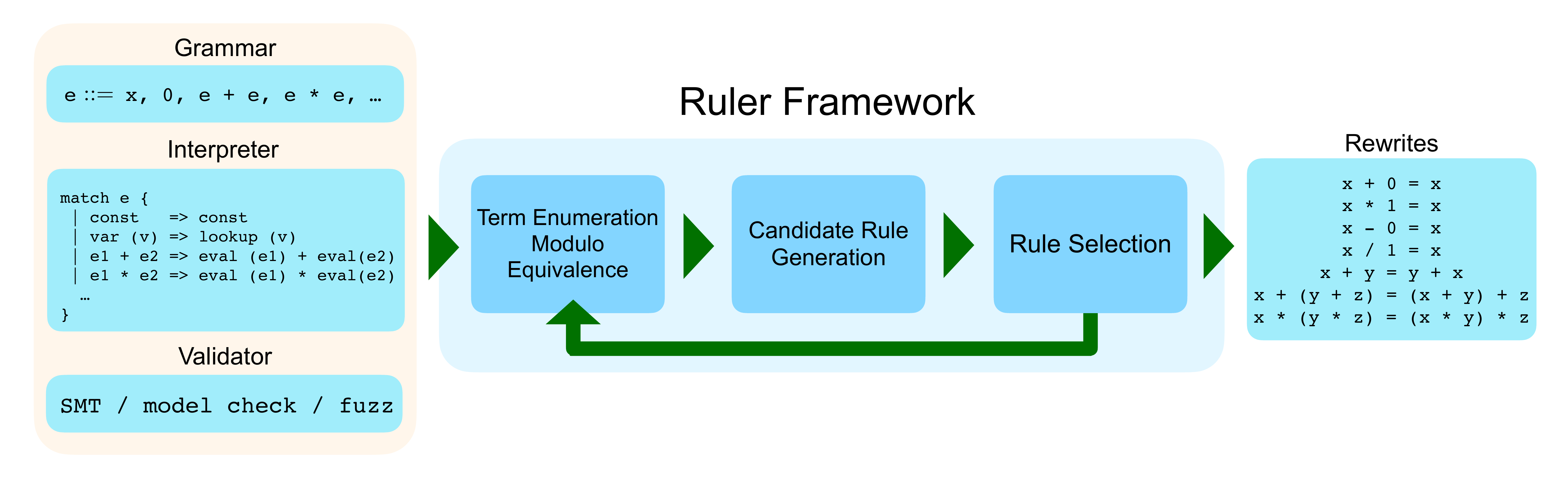}
  \caption{
   \textit{\ruler Workflow}.
    Given a grammar and interpreter for a target domain,
      \ruler uses e-graphs and equality saturation
      to efficiently enumerate potential rewrite rules and
      iteratively select a small set of general, orthogonal rules.
    \ruler supports various validation strategies to
      ensure soundness and speed up synthesis, including
      constraint solving (e.g., SMT), model checking, and fuzzing
      (\autoref{sec:ablation}).
%
%
%
%
 }
 \label{fig:grammar1}
\end{figure*}

%% file: 02-background.tex

\section{Background}
\label{sec:back}





To build a rewrite system for a target domain,
  programmers must develop a set of rewrite rules
  and then use a rewrite engine to apply them,
  e.g., for optimization, synthesis, or verification.
\ruler helps automate this process
  using e-graphs to compactly represent sets of terms and
  using \eqsat to filter and minimize candidate rules.

\subsection{Implementing Rewrite Systems}


\mypara{Developing Rewrite Rules}
Within a given domain $D$,
  a rewrite rule $\ell \rewritesboth r$ is a first-order
  formula consisting of a single equation,
  where $\ell$ and $r$ are terms in $D$
  and all free variables are $\forall$-quantified.
Rewrite rules must be sound:
  for any substitution $\sigma$
  of their free variables,
  $\ell$ and $r$ must have
  the same semantics, i.e.,
  $\llbracket \sigma(\ell) \rrbracket_D =
    \llbracket \sigma(r) \rrbracket_D$.
    
In many cases, rewrite rules must also be engineered to 
  meet (meta)constraints of the rewrite engine responsible for applying them.
For example,
  classic term rewriting approaches often require
  special considerations for 
  cyclic (e.g., $(x + y) \rewritesto (y + x)$) or 
  expansive (e.g., $x \rewritesto (x + 0)$)
  rules~\cite{traat}.
The choice of rules and their ordering can
  also affect the quality and performance of
  the resulting rewrite system~\cite{phase-ordering}.
Different ruleset variations
  may cause a rewrite system to be faster or slower and
  may be able to derive different sets of equivalences.

To a first approximation,
  smaller rulesets of
  more general, less redundant rules are desirable.
Having fewer rules speeds up rule-based search
  since there are fewer patterns to repeatedly match against.
Having more general, orthogonal rules
  also increases a rewrite system's ``proving power''
  by expanding the set of equivalences derivable
  after a smaller number of rule applications.
Avoiding redundancy also aids debugging,
  making it possible to diagnose a misbehaving
  rule-based search or optimization by
  eliminating one rule at a time.

Automatic synthesis aims to generate rulesets that are sound 
  and that include non-obvious, profitable rules
  that even domain experts may overlook for years.\footnote{
    \url{https://github.com/halide/Halide/pull/3719} \\
    \hphantom{2}\url{https://github.com/uwplse/herbie/issues/261} \\
    \hphantom{2}\url{https://github.com/apache/tvm/pull/5974} \\
    \hphantom{2}\url{https://github.com/Z3Prover/z3/issues/2575} \\
    \hphantom{2}\url{https://github.com/Z3Prover/z3/pull/4663}}
Ideally, ruleset synthesis itself should also be fast;
  rapid rule inference can help programmers explore the
  design space for rewrite systems in new domains.
  It can also help with rewrite system maintenance since
  rulesets must be rechecked and potentially extended
  whenever any operator for a domain is
  added, removed, or updated, i.e., when the semantics for the domain evolves.
 


\mypara{Applying Rules with Rewrite Engines}
Given a set of rewrite rules,
  a rewrite engine is tasked with either 
  optimizing a given term into a ``better'' equivalent term
  (e.g., for peepholes~\cite{peephole} or
  superoptimization~\cite{superoptimization})
  or proving two given terms equivalent,
  i.e., solving the \textit{word problem}~\cite{terese}.

Classic term rewriting systems destructively
  update terms as they are rewritten.
This approach is generally fast, but
  complicates support for cyclic or expansive rules,
  and makes both rewriting performance and output quality
  dependent on fine-grained rule orderings.
Past work has extensively investigated
  how to mitigate these challenges by
  scheduling rules~\cite{
    dershowitz1982orderings, knuth1983simple,
    borovansky1998overview, barendregt1987term},
  special casing cyclic and expansive rules~\cite{
    dershowitz1987termination,  bachmair2000congruence, eker2003associative,lucas2001termination}, and
  efficiently implementing rewrite rule-based search~\cite{
    visser2001stratego, visser2001survey, clavel2007all, kirchner2015rewriting}.
Many systems still rely on ad hoc rule orderings
  and heuristic mitigations developed
  through trial and error, though
  recent work~\cite{julie-halide} has demonstrated
  how reduction orders~\cite{traat} can be
  automatically synthesized and then
  used to effectively guide destructive
  term rewriting systems.

\subsection{\Egraphs}
\label{subsec:egg}
\label{subsec:egraphs}


An \textit{equality graph} (\egraph) is a
  data structure commonly used in
  theorem provers~\cite{denali, simplify, nelson}
  to efficiently compute a congruence relation
  over a set of terms. 
\Egraph implementations are backed by 
  union-find data structures~\cite{tarjan}.
An \egraph consists of
  a set of \textit{\eclasses} (equivalence classes) where
  each \eclass contains one or more \textit{\enodes}, and
  each \enode is a tuple of
  a function symbol and a list of children \eclasses.

An \egraph compactly stores an
 equivalence (more specifically, congruence) relation over terms.
We say that an \egraph \textit{represents} the terms 
 in its equivalence relation,
 recursively defined as:
\begin{itemize}
    \item An \egraph represents all terms
      represented by any of its \eclasses.
    \item An \eclass represents all terms
      represented by any of its \enodes.
      Terms represented by the 
        same \eclass are considered equivalent.
    \item An \enode $f (c_1, c_2, ...)$
           represents a term $f (t_1, t_2, ...)$
           iff the function symbols match and each \eclass $c_i$ represents term $t_i$.
         Terms represented by the 
          same \enode are equivalent and have the same top-level function symbol.
\end{itemize}

\Egraphs use hashconsing~\cite{hashconsing} to
  ensure \enodes are never duplicated in an \egraph.
This sharing helps keep \egraphs compact
  even when representing
  exponentially many terms.
\Egraphs may even contain cycles, 
  in which case they represent an infinite set of terms.

\begin{figure}[t]
  \includegraphics[width=70mm]{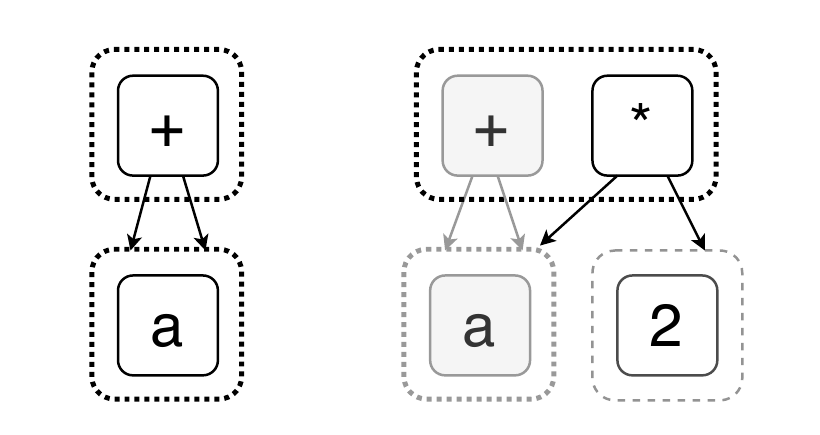}
  \caption{
    Initially, the \egraph (left) represent only
      the term $a + a$.
    The rewrite rule $(x + x) \rewritesboth (x \times 2)$
      is applied to the initial \egraph which
      merges the term $a \times 2$
      into the \textit{same} \eclass that
      contains $a + a$.
    The dotted boxes show \eclasses and
      the solid boxes show \enodes.
    }

  \label{fig:eg-overview}
\end{figure}

\autoref{fig:eg-overview} shows two small \egraphs.
  In the left example,
   each \enode is in its own \eclass.
  This \egraph represents only the term $a + a$.
  In the right example,
   the top \eclass contains two \emph{equivalent} \enodes:
  the \enode with the function symbol $+$
  represents the term $a + a$, and
  the \enode with the function symbol $*$ represents the term $a*2$.

\begin{figure}[t]
\begin{lstlisting}[
  language=Python,
  basicstyle=\normalsize\ttfamily,
  numbers=left,
  xleftmargin=1.5em]
def eq_sat($t$, $R$):
    egraph = make_egraph($t$)
    saturated = false
    while (not saturated) && (not timeout()):
        saturated = true
        for $\ell \rewritesto r$ in $R$:
            for ($\sigma$, $c_\ell$) in egraph.search($\ell$):
                $c_r$ = egraph.add($\sigma$($r$))
                if not egraph.same_eclass($c_\ell$, $c_r$)
                    egraph.union($c_\ell$, $c_r$)
                    saturated = false
    return egraph.extract(cost_fun)
\end{lstlisting}
  \caption{
    Equality Saturation.
    Initially, the \egraph only represents the
    AST of the original input term.
    Semantics-preserving rewrite rules are
    applied until the \egraph saturates or a
    timeout is reached.
    A cost function guides the extraction of the
    ``best'' program from the \egraph.
  }
  \label{fig:eqsat-alg}
\end{figure}

\subsection{Equality Saturation}
\label{subsec:eqsat}

Equality saturation~\cite{eqsat} 
  is a promising technique that 
  repurposes \egraphs to implement
  efficient rewrite engines.
Starting from a term $t$,
  \eqsat builds an initial \egraph $E$
  representing $t$,
  and then repetedly applies rules
  to expand $E$ into a large set
  of equivalent terms.

For optimization or synthesis,
  a user-provided cost function determines the
  ``best'' term $t_b$ also represented by $t$'s \eclass;
  $t_b$ is then extracted and
  returned as the optimized result.
For proving an equivalence $t_1 = t_2$,
  both $t_1$ and $t_2$ are initially added to $E$
  and the \eqsat engine tests whether their
  \eclasses ever merge.

A major advantage of \eqsat engines
  is that they are non-destructive:
  information is never lost when applying a rule.
This enables support for cyclic and expansive rules and
  mitigates many scheduling challenges:
  rule ordering generally does not affect
  the quality of the final result.
Recent work has applied high performance
  \eqsat libraries~\cite{egg} to produce state-of-the-art
  synthesizers and optimizers across several diverse domains~\cite{
    diospyros, tensat, herbie, szalinski,
    spores, yogo, helm}.


\autoref{fig:eqsat-alg} shows
  the core equality saturation algorithm.
Given a term $t$ and a set of rewrite rules $R$,
  the \lstinline{eq_sat} procedure first makes
  an initial \egraph corresponding to $t$'s
  abstract syntax tree (AST);
  each \eclass initially contains a single \enode.
Rewrite rules are then
  repeatedly applied to all terms
  represented in the \egraph.
To apply a rule $\ell \rewritesto r$,
  the algorithm first uses a 
  procedure called \textit{e-matching}~\cite{simplify}
  to search all represented terms
  for those matching $\ell$.
This returns a list of
   pairs ($\sigma$, $c_\ell$)
   of substitutions and the corresponding \eclass
   where the rule matched some term $\sigma(\ell)$.
Each substitution $\sigma$ is then
  \textit{applied} to $r$,
  yielding a new term $\sigma(r)$ equivalent
  to $\sigma(\ell)$.
If $\sigma(r)$ is not already represented by the same \eclass $\sigma(l)$ is, then
$\sigma(r)$ is added 
to a fresh \eclass $c_r$,
  and $c_r$ is unioned (merged) with $c_\ell$,
  the \eclass where $\sigma(\ell)$ was matched.
This process illustrates how
  rewriting in an \egraph is non-destructive:
  applying rules only adds information.


The second \egraph in \autoref{fig:eg-overview}
  shows an example of \egraph rewrite rule application.
Upon applying the rewrite rule
  $(a + a) \rewritesto (a \times 2)$
  to the \egraph on the left,
  a new term is added to the \egraph.
This new term is in the same \eclass
  as the original term,
  indicating they are equivalent. 
One edge from this new \enode points to
  the \eclass containing $x$,
  while the other points to a
  freshly-added \eclass containing
  the \enode for $2$.

When applying rules no longer adds new information
  (i.e., all terms and equalities were already present),
  we say the \egraph has \textit{saturated}.
Upon saturation (or timeout),
  a user-provided cost function
  determines which represented term to
  \textit{extract} from the \egraph.
Equality saturation supports various extraction procedures,
  from simply minimizing AST size
  to more sophisticated procedures based on
  integer linear programming or
  genetic algorithms~\cite{egg}.


%% file: 03-algorithm.tex
\section{Ruler}
\label{sec:core}
\label{sec:meta}
\label{sec:ruler}
\label{sec:over}


This section describes \ruler,
 a new \eqsat-based rewrite rule synthesis technique.
Like other rule synthesis approaches, 
 \ruler iteratively performs three steps:
\begin{enumerate}
    \item Enumerate terms into a set $T$.
    \item Search $T \times T$ for a set of candidate equalities $C$.
    \item Choose a useful, valid subset of $C$ to add to the ruleset $R$.
\end{enumerate}

\ruler's core insight is that \egraphs and \eqsat can help 
 compactly represent the sets $T$, $C$, and $R$, leading 
 to a faster synthesis procedure that produces smaller 
 rulesets $R$ with greater proving power (\autoref{subsubsec:derive}). 
 
\begin{figure}[t]
\begin{lstlisting}[
  basicstyle=\normalsize\ttfamily,
  xleftmargin=2.5em,
  numbers=left]
def ruler (iterations):
  $T$ = empty_egraph()                            $\label{ln:empty-egraph}$
  $R$ = $\{\}$
  for $i \in [0, \texttt{iterations}]\label{ln:iterations}$:
    # add new terms directly to the e-graph representing $T$
    add_terms($T$, $i$)     $\label{ln:add-expressions}$
    loop:$\label{line:inner-loop}$
      # combine e-classes in the e-graph representing $T$ that $R$ proves equivalent
      run_rewrites($T$, $R$)    $\label{ln:run-rewrites}$
      $C$ = cvec_match($T$)             $\label{ln:cvec-match}$
      if $C = \{\}$:  $\label{ln:chk}$
        break
      # choose_eqs only returns valid candidates by using 'is_valid' internally
      # and it filters out all invalid candidates from $C$
      $R$ = $R\ \cup$ choose_eqs($R$, $C$)                 $\label{ln:choose-eqs}$
  return $R$
\end{lstlisting}
  \caption{
  \tool's Core Algorithm.
  The \textsf{iterations} parameter 
  determines the maximum 
  number of connectives in the terms
  \tool will enumerate.
  }
  \label{fig:ruler-core}
\end{figure}
 

\subsection{\ruler Overview}
\label{subsec:ruler-overview}

\autoref{fig:ruler-core} shows \ruler's core
  synthesis algorithm,
  which is parameterized by the following:
\begin{itemize}
    \item The number of iterations to perform the search for (line \ref{ln:iterations});
    
    \item The language grammar, given in the form of a term enumerator 
          (\lstinline{add_terms}, line \ref{ln:add-expressions}),
          which takes the number of variables or constants to enumerate over;
\item The procedure for validating candidate rules, \lstinline{is_valid}
          (called inside \lstinline{choose_eqs}, \autoref{fig:choose-eqs} line \ref{ln:is-valid}).
\end{itemize}

These parameters provide flexibility for
 supporting different domains,
 making \ruler a rule synthesis framework
 rather than a single one-size-fits-all tool. 
Ruler uses an e-graph to compactly represent the set of terms $T$.
In each iteration,
  \ruler first extends the set $T$
  with additional terms from the target language.
Each term $t \in T$ is tagged with a
 \textit{characteristic vector} (cvec) that 
 stores the result of evaluating $t$ given 
 many different assignments of values to variables.

After enumerating terms,
  \ruler uses \eqsat
  (\lstinline{run_rewrites})
  to  merge terms in $T$ 
  that can be proved equivalent 
  by the rewrite rules already discovered (in the set $R$);
Next, \ruler computes a set $C$ of candidate rules (\lstinline{cvec_match}). 
It finds pairs $(t_1, t_2) \in T \times T$
  where $t_1$ and $t_2$
  are from distinct \eclasses
  but have matching \cvecs 
  and thus are likely to be equivalent.
Thanks to \lstinline{run_rewrites},
  no candidate in $C$ should be derivable from $R$.
However, $C$ is often still large and
  contains many redundant or invalid candidate rules.
  
Finally, \ruler's \lstinline{choose_eqs} procedure picks a valid subset of $C$ to add to $R$,
  ideally finding the smallest extension
  which can establish all equivalences implied by $R \cup C$.
\ruler tests candidate rules for validity using a 
 domain-specific \lstinline{is_valid} function.
This process is repeated until there are no
  more equivalences to learn between terms in $T$,
  at which point \ruler begins another iteration.
We detail each of these phases in the rest of this section. 

\subsection{Enumeration Modulo Equivalence}
\label{subsec:por}

Rewrite rules encode equivalences between terms,
  often as relatively small ``find and replace'' patterns.
Thus, a straightforward strategy for
  finding candidate rules is to
  find all equivalent pairs of terms up to some maximum size.
Unfortunately,
  the set of terms up to a given size grows exponentially,
  making complete enumeration impractical for many languages.
This challenge may be mitigated by
  biasing enumeration towards ``interesting'' terms,
  e.g. drawn from important workloads, or by
  avoiding bias and using sampling techniques to
  explore larger, more diverse terms.
\ruler can support both
  domain-specific prioritization and random sampling
  via the \lstinline{add_terms} function.
While these heuristics can be very effective,
  they often risk missing profitable candidates
  for new classes of inputs or use cases.


Term space explosion can also be mitigated by
  partitioning terms into equivalence classes
  and only considering a single, canonical
  representative from each class.
Similar to partial order reduction techniques in
  model checking~\cite{peled1998ten}
  this can make otherwise intractable enumeration feasible.
\ruler defaults to this complete enumeration strategy,
  using an \egraph to compactly represent $T$ and
  \eqsat to keep $T$ partitioned with respect to equivalences
  derivable from the rules in $R$
  even as they are being discovered.
  

 
\mypara{Enumerating Terms in an \Egraph}
\Egraphs are designed to represent
  large sets of terms efficiently by
  exploiting sharing and equivalence.
For sharing, \egraphs maintain deduplication
  and maximal reuse of subexpressions
  via hash-consing.
If some term $a$ is already represented in an \egraph,
  checking membership is constant time and
  adding it again has no effect.
The first time $(a + a)$ is added,
  a new \eclass is introduced with only a single \enode,
  representing the $+$ with
  both operands pointing to $a$'s \eclass.
If $((a + a) * (a + a))$ is then added,
  a new \eclass is introduced with only a single \enode,
  representing the $*$ with
  both operands pointing to $(a + a)$'s \eclass.
Thus as \ruler adds expressions to $T$,
  only the new parts of each added expression
  increase the size of $T$ in memory.

On iteration $i$,
  calling \lstinline{add_terms($T$, i)} 
  adds all (exponentially many) terms
  with $i$ connectives to the \egraph.
The first iteration calls \lstinline{add_terms}
  with an empty \egraph to add terms with $i=0$ connectives,
  thus specifying how many variables and which constants (if any)
  will be included in the search space.
Since these terms are added to an \egraph,
  deduplication and sharing automatically
  provide efficient representation,
  but 
  they do not, by themselves,
  provide an equivalence reduction to help avoid
  enumerating over many equivalent terms.

\mypara{Compacting T using R}
Ruler's \egraph not only stores the set of terms $T$,
 but also an equivalence relation
 (more specifically, a congruence relation)
 over those terms.
Since the children of an \enode are \eclasses, 
 a single \enode can represent 
 exponentially many equivalent terms.
Initially, the \egraph stores no equivalences,
 i.e., each term is in its own equivalence class.
 
As the algorithm proceeds,
 \ruler learns rules and places them in the set $R$ 
 of accepted rules.\footnote{
   While \autoref{fig:ruler-core} shows $R$ starting empty, 
   the user may instead initialize $R$ with trusted axioms if they choose.
 }
At the beginning of its inner loop (line \ref{ln:run-rewrites}),
  \ruler performs \eqsat with the rules from $R$.
Equality saturation will unify classes 
  of terms in the \egraph that can be proven 
  equivalent with rules from $R$.
To ensure that \lstinline{run_rewrites} only
  \emph{shrinks} the term \egraph,
\tool performs this \eqsat
      on a copy of the \egraph,
      and then copies the newly learned equalities
      (\eclass merges)
      back to the original \egraph.
  \change{This avoids polluting the \egraph
    with terms added during \eqsat, 
    i.e.,
    it prevents enumeration of new terms
    based on intermediate terms introduced
    during \eqsat.
    }
 
 
\ruler's inner loop only terminates 
  once there are no more rules to learn,
  so the next iteration
  (\lstinline{add_terms}, line \ref{ln:add-expressions})
  \textit{only enumerates over the canonical representatives}
  from the equivalence classes of terms with respect to $R$
  that have been represented up to that point.
This compaction of the term space makes
  complete enumeration possible for non-trivial depths
  and makes \ruler much more efficient in
  finding a small set of powerful rules.
\autoref{sec:ablation} demonstrates 
 how compaction of $T$ is essential to \ruler's performance.
 
Since $R$ may contain rules that use partial operators,
 \ruler's \eqsat implementation 
 only merges \eclasses whose \cvecs 
 agree in at least one non-null way
 (see the definition of \textit{match} in \autoref{subsec:cvecmatch}).
For example, consider that $x/x \rewritesboth 1 \in R$, 
 and both $\frac{a+a}{a+a}$ and $\frac{a-a}{a-a} \in T$.
The pattern $x/x$ matches both terms, but
 \eqsat will not merge $\frac{a-a}{a-a}$ with $1$,
 since $\frac{a-a}{a-a}$ is never defined.
On the other hand,
 $\frac{a+a}{a+a}$ can merge with $1$
 since the \cvecs match.

Prior work~\cite{sat19} on rule inference
  applies multiple filtering passes to minimize rule sets
  \textit{after} they are generated.
These filters include
  subsumption order,
  variable ordering,
  filtering modulo alpha-renaming,
  and removing rules in the congruence closure
  of previously found rules.
\ruler eliminates the need for such filtering
  using \eqsat on the \egraph representing $T$.
Since enumeration takes place over \eclasses in $T$,
 equivalent terms are ``pre-filtered'' automatically.
 
\subsection{Candidate Rules}
\label{subsec:cands}

Given a set (or in \ruler's case, an \egraph) of terms $T$,
 rewrite rule synthesis
 searches $T \times T$ for pairs of equivalent terms that
 could potentially be a rule to add to $R$.
The set of candidate rules is denoted $C$.
 
The naive procedure for producing candidate rules
 simply considers every distinct pair: 
 $$C = \{ l \rewritesboth r \mid l,r \in T.\ l \not= r \wedge \forall \sigma.\ l[\sigma] = r[\sigma] \}$$
This is prohibitively expensive for two main reasons.
First, it will produce many rules that 
 are either in or can be proven by the existing ruleset $R$.
In fact, the naive approach should always produce
 supersets of $C$ from previous iterations;
 accepting a candidate rule from $C$ into $R$ 
 would not prevent it from being generated in $R$ in the following iteration.
Second, most of the candidates will be unsound,
 and sending too many unsound candidates to \lstinline{choose_eqs}
 burdens it unnecessarily,
 since it must search $C$ for valid candidates by 
 invoking the user-supplied \lstinline{is_valid} procedure.
\ruler's use of an \egraph to represent the term set $T$
 addresses both of the these inefficiencies 
 with techniques called 
 \textit{canonical representation}
 and
 \textit{characteristic vectors}.
 
\mypara{Canonical Representation}
Consider a situation where
 $(x + y) \rewritesboth (y + x) \in R$
 and both  
 $(a + b)$ and $(b + a)$ are in $T$.
When selecting terms from which to build a candidate rule,  
 considering both 
 $(a + b)$ and $(b + a)$
 would be redundant; 
 any rules derived from one could be derived from the other 
 by composing it with commutativity of $+$.
In some rewriting systems, 
 this composition of rewrites cannot be achieved
 since cyclic rules like commutativity are not permitted.
Equality saturation, however,
 handles and in many cases prefers such compositional rules,
 since it results in fewer rules to search over the \egraph.
 
To prevent generating candidate rules which are already
 derivable by the rules in $R$,
 \ruler only considers a single term from each 
 \eclass when building candidate rules.
 When searching for candidate rules, 
 \ruler considers only term pairs $(l, r)$ 
 where $l \neq r$ and both are canonical representatives 
 of \eclasses in $T$.
This ensures candidate rules cannot be derived from $R$; 
 if they could have been,
 then $l$ and $r$ would have been in the same \eclass
 after the call to \lstinline{run_rewrites}.

\mypara{Characteristic Vectors}
\label{subsec:cvecmatch}
%

Canonical representation reduces
 $C$ from $T \times T$ to $T' \times T'$
 where $T'$ is the set of canonical terms from $T$,
 but it does not prevent a full $O(n^2)$
 search of $T' \times T'$ for valid candidate rules.
Ruler employs a technique called 
 \textit{characteristic vectors} (\cvecs) to
 prevent this quadratic search by only considering 
 pairs that are \textit{likely} valid.
Ruler associates a characteristic vector $v_i$ 
 with each \eclass $i$.
The \cvec is the result of evaluating
 $t_i$, the canonical term in \eclass $i$,
 over a set of inputs
 that serves as a ``fingerprint''\footnote{
   \autoref{sec:related}
   discusses prior work~\cite{taso19, bansal} that uses
   ``fingerprints" for 
   synthesizing peephole optimizations and graph
   substitutions.}
 for the value of that \eclass.
Stated precisely, 
 let 
  $\sigma_j$ for $j \in [1, m]$ be a predetermined family of $m$
  mappings from variables in $T$ to concrete values,
  and let \textsf{eval} be the evaluator for the given language.
The \cvec for \eclass $i$ is:
 $$v_i = [ \textsf{eval}(\sigma_j, t_i) \mid j \in [1, m] ]$$
 
\ruler computes \cvecs
 incrementally and without redundancy during enumeration
 using an \textit{\eclass analysis} \cite{egg}
 to associate a \cvec with each \eclass;
 let $i$ be an \eclass, $t_i$ its canonical term, and $v_i$ its cvec:
\begin{itemize}
    \item 
  when $t_i = n$ for a constant $n$,
    $v_i$ is populated by copies of $n$;
    \item 
  when $t_i = f(t_{j_1}, \ldots, t_{j_n})$ for some $n$-ary operator $f$ from the given language,
    $v_i$ is computed by mapping $f$ over
    the \cvecs of the subterms:\;
    $v_i = \textsf{map}(f, \textsf{zip}(v_{j_1}, \ldots, v_{j_n}))$
    \item 
  when $t_i = x$ for a variable $x$,
    $v_i$ is populated by values from the target domain;
  choosing values to populate the \cvecs of variables can be done randomly 
  or with a domain-specific approach
  (\autoref{sec:ablation} compares two approaches).
\end{itemize}

To support partial operators (e.g., division), 
 \cvecs may have a null value in them to indicate failure to evaluate.
We say that \cvecs \textit{match} if their 
 non-null values agree in every (and at least one) position,
 i.e., \cvecs $ [a_1, \ldots, a_n]$ and $[b_1, \ldots, b_n]$ match iff:
$$
 \forall i.\ a_i = b_i \vee a_i = \textsf{null} \vee b_i = \textsf{null} 
 \quad\textrm{  and  }\quad
 \exists i.\ a_i = b_i \wedge a_i \neq \textsf{null} \wedge b_i \neq \textsf{null} 
$$
 
When \eclasses in the \egraph representing $T$ merge, 
 they will have matching \cvecs, 
 because they have been proven equivalent by valid rules.
Ruler aborts if \cvecs of merging \eclasses do not match;
 empirically, this helps avoid learning unsound rules 
 even when \lstinline{is_valid} is not sound (\autoref{subsec:soundness}).
 
\autoref{sec:eval} and \autoref{sec:case} discuss
  how \cvecs are generated for different domains.
Characteristic vectors serve as a filter for validity:
 if $i,j$ are \eclasses and $v_i$ does not match $v_j$,
 (using the definition of match from \autoref{subsec:ruler-overview})
 then $t_i \rewritesboth t_j$ is not valid.
This allows \ruler to not consider 
 those pairs when building $C$:
$$ C = 
\{ t_i \rewritesboth t_j \mid 
   i,j \in \text{\eclasses of $T$}.\ \textsf{match}(v_i, v_j) \}
$$


The 
 \lstinline{cvec_match} procedure
 (called at \autoref{fig:ruler-core},
  line \ref{ln:cvec-match})
 constructs $C$ by
 grouping \eclasses from $T$
 based on their \cvecs
 and then taking pairs of canonical terms 
 from each of those groups.

\mypara{Validation}
The candidate set $C$ contains rules that are likely,
 but not guaranteed, to be valid.
The \lstinline{choose_eqs} function 
 (discussed in \autoref{subsec:choose})
 must validate these before returning them by using the
 user-supplied \lstinline{is_valid} function.
The soundness of \ruler's output, i.e., 
 whether every rule in $R$ is valid,
 depends on the soundness of the provided 
 \lstinline{is_valid} procedure.
Many rule synthesis implementations \cite{swapper, taso19}
 use SMT solvers to perform this validation.
Ruler supports arbitrary validation procedures:
 small domains may use model checking,
 larger domains may use SMT, 
 and undecidable domains may decide to give up 
 a guarantee of soundness and use a sampling-based validation.
\autoref{subsec:soundness} 
 compares validation 
 techniques for different domains.
 
 
\subsection{Choosing Rules}
\label{subsec:select}
\label{subsec:choose}

\begin{figure}[t]
\begin{lstlisting} [
 numbers=left,
 basicstyle=\footnotesize\ttfamily,
 xleftmargin=7mm,
]
# $R$ is the accepted ruleset so far, $C$ is the candidate ruleset.
# Ruler's implementation of choose_eqs is based on a more flexible choose_eqs_n.
def choose_eqs($R$, $C$, $n = \infty\label{line:choose-eqs-n}$):
   for $\mathit{step} \in [100,\, 10,\, 1]\label{line:step}$:
       if $\mathit{step} \leq n$:
           $C$ = choose_eqs_n($R$, $C$, $n$, $\mathit{step}$)
   return $C$
   
# $n$ is the number of rules to choose from $C$, and $\mathit{step}$ is a granularity parameter.
# A larger $\mathit{step}$ size allows you to eliminate redundant rules faster.
def choose_eqs_n($R$, $C$, $n$, $\mathit{step}$):
   # let $K$ be the list of "keepers" which we will return
   $K = []$
   while $C \neq \emptyset\label{line:choose-eqs-loop}$: 
      # pick the best $\mathit{step}$ candidate rules from $C$ according to a heuristic 
      # that approximates rule "generality", including subsumption.
      $C_{\sf best},\; C$ = select($\mathit{step}$, $C$)
      
      # add the valid ones to $K$
      $K$ = $K \cup \{c \mid c \in C_{\sf best}.\ \texttt{is\_valid}(c) \}\label{ln:is-valid}$
      
      # remember all the invalid candidates in a global variable $\mathit{bad}$;
      # Ruler uses this to prevent known-invalid candidates from entering $C$ again (not shown)
      $\mathit{bad}$ = $\mathit{bad} \cup \{c \mid c \in C_{\sf best}.\ \neg\texttt{is\_valid}(c) \}$
      
      # stop if we have enough rules
      if $|K| \geq n$: $\label{line:stop-n}$
         return $K[0..n]$
            
      # try to prove terms remaining in $C$ equivalent using rules from $R \cup K$
      $C$ = shrink($R \cup K$, $C$)
   return $K$
   
def shrink($R$, $C\label{line:shrink}$):
   $E$ = empty_egraph()
   for $(l \rewritesboth r) \in C$:
      $E$ = add_term($E$, $l$)
      $E$ = add_term($E$, $r$)
   $E$ = run_rewrites($E$, $R$)
   # return the extracted versions of rules from $C$, leaving out anything that was proven equivalent
   return $\{\texttt{extract}(E, l) \rewritesboth \texttt{extract}(E, r)
            \mid  (l \rewritesboth r) \in C.\;
            \neg \texttt{equiv}(E, l, r)
            \}$
               
\end{lstlisting}
  \caption{
    \ruler's implementation of \lstinline{choose\_eqs},
    which aims to minimize the candidate set $C$
    by eliminating subsets that the remainder can derive.
  }
  \label{fig:choose-eqs}
\end{figure}

After finding a set of candidate rules $C$,
  \ruler selects a valid subset of rules from $C$ 
  to add to the rule set $R$
  using the \lstinline{choose_eqs} procedure
  (\autoref{fig:ruler-core}, line \ref{ln:choose-eqs}).
As long as \lstinline{choose_eqs} returns 
 a valid, non-empty subset of $C$,
 \ruler's inner loop will terminate:
 the number of \eclasses with matching \cvecs
 (i.e., the subset of $T$ used to compute $C$)
 decreases in each iteration since
 $R$ is repeatedly extended with
 rules that will cause new merges
 in \lstinline{run_rewrites}.
Ideally, \lstinline{choose_eqs}
  quickly finds a minimal extension of $R$
  that enables deriving all equivalences
  implied by $R \cup C'$ where $C'$ is 
  the valid subset of $C$.
\change{
  \lstinline{choose_eqs} also removes
  invalid candidates
  from $C$;
  if it returns the empty set
  (i.e., none of the candidates in $C$ are valid),
  then \ruler's inner loop will terminate in the next iteration
  due to line \ref{ln:chk} in \autoref{fig:ruler-core}.}

  

The candidate rules in $C$ are 
 not derivable by $R$, 
 but many of the 
 candidate rules may be able to derive each other,
 especially in the context of $R$.
For example, the following candidate set
 is composed of three rules from the boolean domain,\footnote{\lstinline{^} here represents XOR}
 and any two can derive the third:
 
\noindent
\hfill
\lstinline{(^ x x) = false} 
\hfill
\lstinline{(& x false) = false}
\hfill
\lstinline{(& x false) = (^ x x)}
\hspace{5em}

An implementation of \lstinline{choose_eqs} 
 that only returns a single rule $c \in C$
 avoids this issue,
 since adding $c$ to $R$ prevents
 those rules derivable by $R \cup \{c\}$ from being 
 candidates in the next iteration of the inner loop.
However, 
 a single-rule implementation
 will be slow to learn rules,
 since it can only learn one at a time 
 (\autoref{tab:eval} of our evaluation shows there are sometimes thousands of rules to learn).
Additionally, 
 such an implementation has to decide which rule to select,
 ideally picking the ``strongest'' rules first.
For example,
 if $a,b \in C$ and $R \cup \{a\}$
 can derive $b$ but
 $R \cup \{b\}$
 can not derive $a$,
 then selecting $b$ before $a$ would be a mistake,
 causing the algorithm to incur an additional loop.
 
  
\ruler's implementation of
 \lstinline{choose_eqs}, shown in \autoref{fig:choose-eqs},
 is parameterized by a value $n$ with default of $\infty$. 
At $n=1$, \lstinline{choose_eqs} simply returns a single valid candidate from $C$.
For higher $n$, \lstinline{choose_eqs} 
 attempts to return a list of up to $n$ valid rules all at once.
This can speed up \ruler by requiring fewer trips around its inner loop,
 but risks returning many rules that can derive each other.
To mitigate this, \lstinline{choose_eqs} tries to not 
 choose rules that can derive each other.
In its main loop (line \autoref{line:choose-eqs-loop}),
 \lstinline{choose_eqs} 
 uses the \lstinline{select} function to pick the $\mathit{step}$ 
 best rules from $C$ according to a syntactic heuristic.\footnote{
 \Ruler's syntactic heuristic prefers candidates 
  with the following characteristics (lexicographically):
 more distinct variables,
 fewer constants,
 shorter larger side (between the two terms forming the candidate),
 shorter smaller side,
 and fewer distinct operators.
}
\Ruler then validates the selected rules
 and adds them to a set $K$ of ``keeper'' rules which it will ultimately return.
It then employs the 
 \lstinline{shrink} procedure (line \autoref{line:shrink})
 to eliminate candidates from $C$ that can be derived be $R \cup K$.
This works similarly to
 \lstinline{run_rewrites} in the \ruler algorithm, 
 but \lstinline{shrink} works over the remaining
 \textit{candidate set} $C$ instead of $T$.

\ruler's \lstinline{choose_eqs} invokes the inner 
 \lstinline{choose_eqs_n} procedure with increasingly small step sizes
 ($\mathit{step}$ is defined on line \ref{line:step}).
Larger step sizes allow \lstinline{shrink} to quickly 
 ``trim down'' $C$ when it contains many candidates.
However, a large step also means that 
 \lstinline{choose_eqs} may admit $\mathit{step}$ rules into $K$ at once, 
 some of which may be able to prove each other.
Decreasing the step size to 1 eliminates this issue.

\ruler uses $n=\infty$ by default for maximum performance, and
 \autoref{sec:ablation} measures the effects of this choice
 on \ruler's performance and output.

\subsection{Implementation}

We implemented \tool in Rust
 using the \egg~\cite{egg} \egraph library for
 equality saturation.
By default \ruler uses Z3~\cite{z3} for SMT-based validation,
 although using other validation backends
 is simple (\autoref{sec:ablation}).
 
\ruler's core consists of under 1,000 lines of code,
 allowing it to be simple, extensible,
 and generic over domains.
Compared to the rewrite synthesis tool
 inside the CVC4 solver~\cite{cvc4, sat19},
 \ruler is an order of magnitude smaller.
 Since \tool's core algorithm does not rely on SMT,
 \tool can learn rewrite rules over domains 
 unsupported by SMT-LIB~\cite{smtlib},
 or for alternative semantics for those domains
 \footnote{For example, the Halide~\cite{halide} tool uses division semantics where $x/0=0$; 
  this is different from the SMT-LIB semantics, but it can easily be encoded using the \texttt{ite} operator.}
 (\autoref{subsec:updates}).

In the following sections, we provide various evaluations of
  three representative domains on top of \ruler's core.
Each domain highlights a
  verification back-end and \cvec generation strategy
  \tool supports:
 \begin{itemize}
 \item \booleans and \bfour: these are small domains which
   \tool can efficiently model check and generate sound rules by construction ---
   the \cvecs are complete.
 \item \bthreetwo: demonstrates that \tool supports SMT-based
   verification for large, non-uniform domains.
 \item \rationals: demonstrates that random sampling
   is adequate for larger but continuous domains.
   This domain also showcases \tool's support for partial operators like division.
 \end{itemize}
 
The implementation of booleans, bitvectors, and rationals are
 in approximately 100, 400, and 300 lines, respectively.


 

%% file: 04-evaluation.tex

\section{Evaluation}
\label{sec:eval}

In evaluating \tool, we are interested
  in the following research questions:
  
\begin{itemize}
    \item \textit{Performance}. 
    \change{
      Does \ruler synthesize rewrite rules quickly compared to similar approaches?
      }
    \item \textit{Compactness}. 
      Does \ruler synthesize small rulesets?
    \item \textit{Derivability}.
    \change{
      Do \ruler's rulesets derive the rules produced by similar approaches?
      }
    \item \textit{End-to-End}.
    \change{
      How well do \ruler's synthesized rules perform compared to 
      rules generated by experts in real applications?
      }
    \item \textit{Sensitivity Analysis}.
    \change{
      How do the different components of \ruler's core algorithm
      affect performance and the size of the synthesized ruleset?
      }
    \item \textit{Validation Analysis}.
      How do different validation strategies affect \tool's output?
\end{itemize}

We first evaluate performance, compactness,
 and derivability (\autoref{subsec:cvc4})
 by comparing \tool against recent work \cite{sat19}
 that added 
 rewrite rule synthesis
 to the CVC4 SMT solver \cite{cvc4}.
We compose \tool's rules with Herbie~\cite{herbie}
  to show an end-to-end evaluation (\autoref{sec:herbie}).
We then study the effects of different choices in \tool's
 search algorithm
 and the different validation and \cvec generation strategies
 (\autoref{sec:ablation}).

\subsection{Comparison with CVC4}
\label{subsec:cvc4}
Both Ruler and the CVC4 synthesizer
 are written in systems programming languages 
 (Rust and C++, respectively),
 and both take a similar approach to 
 synthesizing rewrite rules:
  enumerate terms,
  find valid candidates,
  select rules, and repeat. 
  
At the developers' suggestion, 
 we used CVC4 version 1.8
 with 
 \textsf{--sygus-rr-synth} to synthesize rules.
We enabled
 their rule filtering
 techniques
 (\textsf{--sygus-rr-synth-filter-cong},
 \textsf{--sygus-rr-synth-filter-match},
 \textsf{--sygus-rr-synth-filter-order}).
We enabled their rule checker
 (\textsf{--sygus-rr-synth-check}) to
 verify all synthesized rules.
 Additionally, we also disabled use of any pre-existing rules from
 CVC4 to guide the rule synthesis
 (using \textsf{--no-sygus-sym-break}, \textsf{--no-sygus-sym-break-dynamic}).
 
\autoref{tab:eval} shows the results of the comparison.
The following text discusses the results in detail, 
 but, in short,
 \ruler synthesizes smaller rulesets in less time that have more proving power (\autoref{subsubsec:derive}).

\begin{table*}
  \centering
  \begin{tabular}{lr|rrl|rrl|rr}
        \multicolumn{2}{c}{Parameters} &
        \multicolumn{3}{c}{Ruler} &
        \multicolumn{3}{c}{CVC4} &
        \multicolumn{2}{c}{Ruler / CVC4} 
        \\
        Domain & \# Conn &
        Time (s) & \# Rules & Drv &
        Time (s) & \# Rules & Drv &
        Time & Rules
        \\
        \hline
bool & 2 & 0.01 & 20 & 1 & 0.13 & 53 & 1 & 0.06 & 0.38\\
bool & 3 & 0.06 & 28 & 1 & 0.82 & 293 & 1 & 0.07 & 0.10\\
bv4  & 2 & 0.14 & 49 & 1 & 4.47 & 135 & 0.98 & 0.03 & 0.36\\
bv4  & 3 & 4.30 & 272 & 1 & 372.26 & 1978 & 1 & 0.01 & 0.14\\
bv32 & 2 & 13.00 & 46 & 0.97 & 18.53 & 126 & 0.93 & 0.70 & 0.37\\
bv32 & 3 & 630.09 & 188 & 0.98 & 1199.53 & 1782 & 0.91 & 0.53 & 0.11\\
\hline
\multicolumn{8}{r|}{} & 0.04 & 0.17 \\
\multicolumn{8}{r}{} & \multicolumn{2}{r}{Harmonic Mean}
    \end{tabular}
    \vspace{1em}
  \caption{
    \ruler tends to synthesize smaller, more powerful rulesets in less
     time than CVC4.
    The table shows synthesis results
     across domains
     and maximum term size (in number of connectives, ``\# Conn'').
    \change{The domains are \booleans, \bfour, and \bthreetwo each of which is
    evaluated for \# Conn = 2 and \# Conn = 3.} 
    For verification, \ruler uses model checking for \booleans and \bfour
     and Z3 for \bthreetwo.
    The ``Drv'' column shows the fraction 
     that tool's synthesized ruleset can derive of the other's ruleset;
     for example, the final row indicates that
     \ruler's 188 rules derived 98\% of CVC4's 1,782 rules,
     and CVC's rules derived 91\% of \ruler's.
    The final two columns show the ratios of 
     synthesis times and ruleset sizes between the two tools.
  }
  \label{tab:eval}
\end{table*}

\begin{figure}
    \centering
    \begin{subfigure}[t]{0.4\linewidth}
    \begin{grammar}
    <var> ::= x | y | z
    
    <expr> ::= <literal> | <var> 
      \alt $\sim$ <expr> 
      \alt <expr> $\&$ <expr> 
      \alt <expr> \texttt{\^} <expr> 
      \alt (<expr> $\mid$ <expr>)
    \end{grammar}
    \caption{Boolean (bool) grammar.}
    \end{subfigure}
    \hfill
    \begin{subfigure}[t]{0.55\linewidth}
    \begin{grammar}
    <var> ::= x | y | z
    
    <expr> ::= <literal> | <var>
      \alt $\sim$ <expr> | $-$ <expr> | 
      \alt <expr> $+$ <expr> | <expr> $-$ <expr> | <expr> $\times$ <expr>
      \alt <expr> $\ll$ <expr> | <expr> $\gg$ <expr>     
      \alt <expr> $\&$ <expr> | (<expr> $\mid$ <expr>)
    \end{grammar}
    \caption{
    Bitvector grammar used for both bv4 and bv32.
    $\sim$ is bitwise negation; unary $-$ is two's-complement.
    We use shift semantics where $a \ll b = a \gg b = 0$ when $b \geq \mathit{width}$.
    }
    \end{subfigure}
    \caption{
      The grammars used by both \ruler and CVC4 in our evaluation.
    }
    \label{fig:bv4-bool-grammar}
\end{figure}

%

\subsubsection{Benchmark Suite}

We compare \ruler against CVC4
 for \booleans, \bfour, and \bthreetwo.
\autoref{fig:bv4-bool-grammar} shows the grammars.
Both \ruler and CVC4
 are parameterized by the domain 
 (bool, bv4, or bv32),
 the number of distinct variables in the grammar,
 and the size of the synthesized term.\footnote{
   Size is measured in number of connectives, 
   e.g., $a$ has 0, $(a + b)$ has 1, 
   and $(a + (b + c))$ has 2.
   In CVC4, this is set with the 
   \textsf{--sygus-abort-size} flag.
 }
All benchmarks were single-threaded 
 and run on an
 AMD 3900X 3.6GHz processor with 32GB of RAM.
Both \ruler and CVC4 were given 3 variables 
 and no constants to start the enumeration.


\subsubsection{Derivability}
\label{subsubsec:derive}

A bigger ruleset is not necessarily a better ruleset.
We designed \ruler to minimize ruleset size 
 while not compromising on
 its capability to prove equalities.
We define a metric called the \textit{deriving ratio}
 to compare two rulesets.
Ruleset $A$ has deriving ratio $p$ with respect to ruleset $B$
 if set $A$ can derive a fraction $p$ of the rules in $B$
 ($A \vDash b$ means rule set $A$ can prove rule $b$):
$$p = |B_A| / |B| \quad\textrm{ where }\quad B_A = \{b \mid b \in B.\ A \vDash b\}$$
If $A$ and $B$ have deriving ratio of 1 with respect to each other,
  then they can each derive all of the other's rules.

We use \egg's equality saturation procedure to test derivability.
To test whether $A \vDash b$ (where $b = b_l \rewritesboth b_r$)
 we add $b_l$ and $b_r$ to an empty \egraph,
 run \eqsat using $A$, 
 and check to see if the \eclasses of $b_l$ and $b_r$
 merged.
We run \egg with 5 iterations of equality saturation.
Since this style of proof is bidirectional 
 (\egg is trying to rewrite both sides at the same time),
 derivations of $b_l = b_r$ can be as long as 10
 rules from $A$.

\subsubsection{Bitvector and Boolean Implementation}

\ruler
 supports different implementations of the \textsf{is_valid} procedure
 (\autoref{subsec:cands}) for different domains.
When the domain is small enough,
 \ruler can use efficient model checking.
For example, there are only $(2^4)^3 = 4096$ 
 assignments of \bfour to three variables.
By using \cvecs of that length to capture all possibilities, 
 \ruler can guarantee that the 
 \lstinline{cvec_match} procedure returns only valid
 candidate rules,
 and \lstinline{is_valid} need not perform any additional checking.
\ruler uses model checking for \booleans and \bfour,
 and it uses SMT-backed verification for \bthreetwo.
 
\subsubsection{Results}

\autoref{tab:eval} shows the results of our comparison with CVC4's
 rewrite rule synthesis.
On average (harmonic mean), 
 \ruler produces $5.8\times$ smaller 
 rulesets $25\times$ faster than CVC4.
Ruler and CVC4's results can derive most of each other.
On the harder benchmarks (in terms of synthesis times), 
 \ruler's results have a higher derivability ratio;
 they can prove more of CVC4 rules than vice-versa.

%% file: 05-casestudy.tex
\section{End-To-End Evaluation of \tool: A Rational Case Study}
\label{sec:herbie}
\label{sec:case}
\label{sec:end2end}

How good are \tool's rewrite rules?
Can they be used with existing rewrite-based tools with little additional effort?
This section demonstrates how
  \tool's output 
  can be plugged directly into
  an existing rewrite-driven synthesis tool, Herbie.
  
    

\begin{figure}
\begin{subfigure}[t]{0.3\linewidth}
   \includegraphics[width=\linewidth]{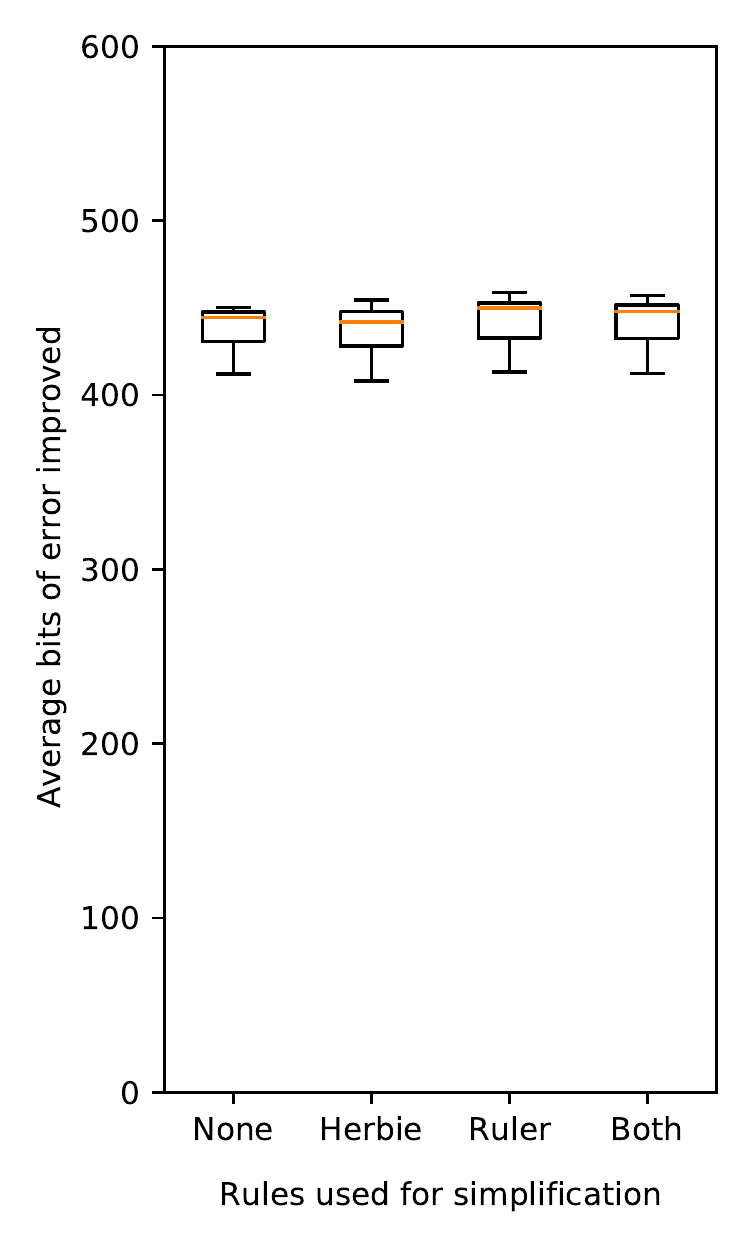}
   \caption{
     Improvement in average error,
     Herbie's metric for measuring accuracy
     (higher is better).
    }
\end{subfigure}
\hfill
\begin{subfigure}[t]{0.3\linewidth}
   \includegraphics[width=\linewidth]{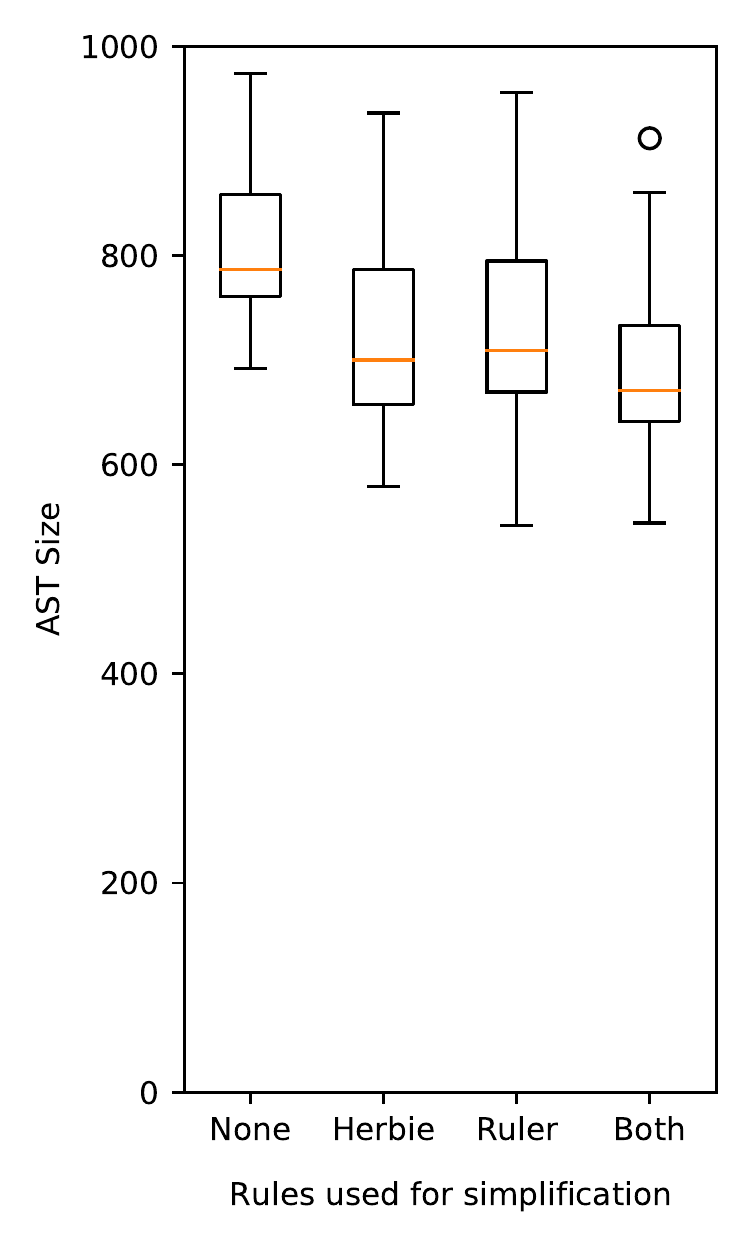}
   \caption{
     Size of the output AST produced by Herbie
     (lower is better).
   }
\end{subfigure}
\hfill
\begin{subfigure}[t]{0.3\linewidth}
  \includegraphics[width=\linewidth]{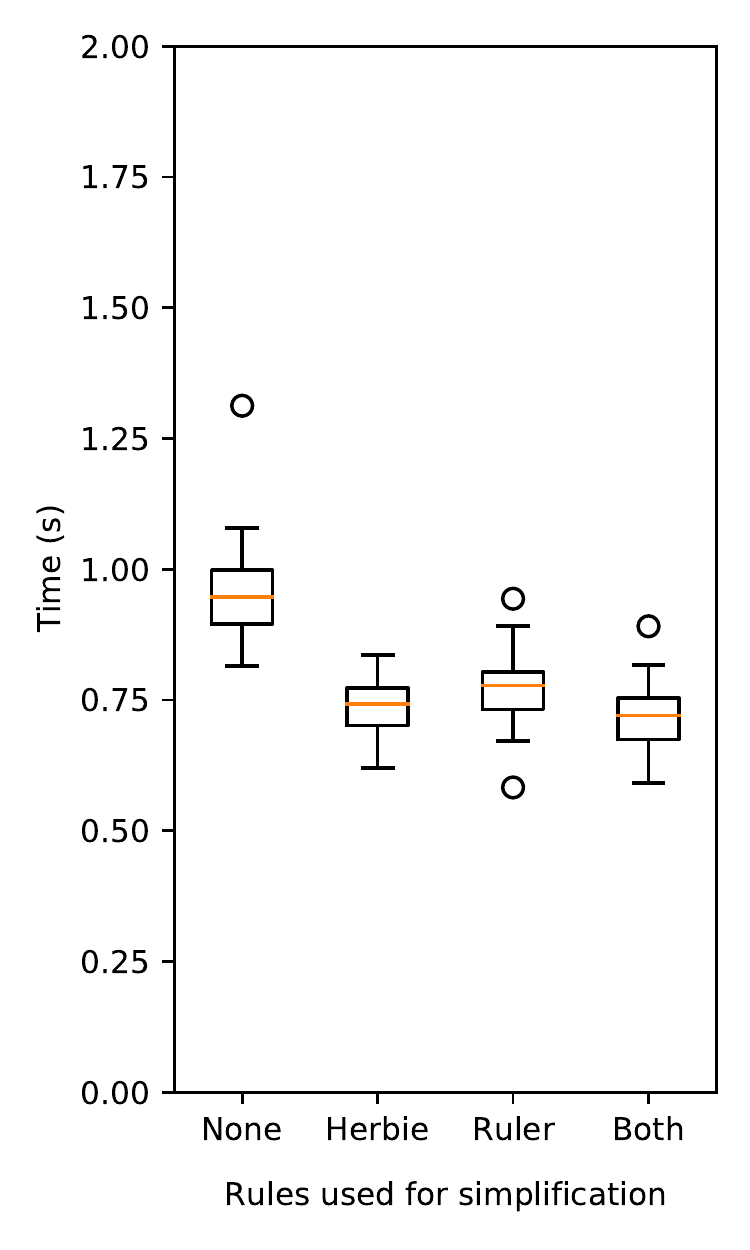}
  \caption{
  Herbie's running time (lower is better).
  }
\end{subfigure}
\caption{ 
 Comparing Herbie results between four configurations.
Each boxplot represents the results from 30 seeds,
 where each data point is obtained by summing the value
 (average error, AST size, time) over all 51 benchmarks.
The columns dictate what rational rules Herbie has access to:
 either none, its default rules, only \ruler's rules, or both.
Herbie's rational rules reduce AST size 
 and speed up simplification without reducing accuracy,
 and \ruler's rules perform similarly (with or without Herbie's rules).
}
\label{fig:herbie-res}
\label{fig:herbie2}
\end{figure}

Herbie~\cite{herbie} is a widely-used, open-source tool
  for automatically improving the accuracy of floating point expressions,
  with thousands of users and regular yearly releases.
Herbie takes as input a numerical expression and returns
  a more accurate expression.
It is implemented in Racket~\cite{racket}.
Herbie has separate phases for error localization (by sampling),
  series expansion, regime inference, and simplification,
  which work together to increase accuracy of numerical programs.
The simplification phase uses 
  algebraic rewrites to simplify mathematical expressions,
  thereby also enabling further accuracy improvements.
These are applied using an equality saturation engine.
In the past, the set of algebraic rewrites has been the cause of many bugs;
  of 8 open bugs at the time of this writing, six have been tagged ``rules'' by the developers~\cite{herbie-issues}.
\ruler was able to find rules that addressed
 one of these issues.

\subsection{Experimental Setup}

We implemented rationals in \tool
 using rational and bigint libraries in Rust~\cite{rust-bigint, rust-rat}.
We then synthesized rewrite rules
  over rational arithmetic and ran Herbie
  with the resulting ruleset.
  
The Herbie benchmark suite has 155 benchmarks;
  55 of those are over rationals --- i.e., all expressions in
  these benchmarks consist only of operators:   $+, -, \times, /, abs, neg$.
At the developers' suggestion,
  we filtered out 4 of the 55 benchmarks because they repeatedly timed out.
We ran all our experiments on the remaining 51 benchmarks under four different configurations:
\begin{itemize}
\item \lstinline{None}: remove all the
  rational rewrite rules from Herbie's simplification phase.
Rational rules are those which consist only of rational operators
  and no others.
Note that all other components of Herbie are left intact,
  including rules over
  rational operators combined with other operators,
  and rules entirely over
  other operators.
\lstinline{None} is the baseline.
\item \lstinline{Herbie}: no changes to Herbie, simply run it on the 51 benchmarks.
\item \lstinline{Ruler}: replace Herbie's rational rules with output of \tool.
\item \lstinline{Both}: run Herbie with both \tool's rational rules and the
  original Herbie rational rules.
  \end{itemize}
  
Herbie has a default timeout of 180 seconds for each benchmark.
It has a node limit of 5000 in its underlying
  equality saturation engine, i.e., it stops applying the simplification rules
  once the \egraph has 5000 \enodes.
We ran our experiments with three settings --- (1) using the defaults,
(2) increasing the timeout to 1000 seconds, and the node-limit to 10,000 to
  account for the addition of extra rules to Herbie's ruleset, and
(3) decreasing the node limit to 2500 --- we found that our results were
  stable and robust across all three settings.
 \autoref{fig:herbie-res} shows the results for the default setting.
For all four configuration
(\lstinline{None, Herbie, Ruler, Both}),
we ran Herbie for 30 seeds (because Herbie's error localization
relies on random sampling).

We used \ruler to synthesize rational rules of depth 2 with 3 variables
  using random testing for validation (``rational" under \autoref{tab:valid-ablation}).\footnote{
  For rationals, the \lstinline{add_terms} implementation enumerates terms by depth rather
  than number connectives, since that matches the structure of Herbie's 
  existing rules.
}
\tool learned 50 rules in 18 seconds, all of which were proven sound with an SMT post-pass.
Four rules were expansive --- i.e., rules like $( a \rewritesboth (a \times 1) )$ whose LHS is only a variable.
We removed these expansive rules from the ruleset as per the recommendation of
  the Herbie developers.
Herbie's rules are uni-directional --- we therefore expanded our
 rules for compatibility, ultimately leading to 76 uni-directional
 \tool rules.

\subsection{Discussion}

The Herbie simplifier uses equality saturation
  to find smaller, equivalent programs.
The simplifier itself does not directly improve accuracy;
  rather, it generates more candidates that are then used in the other
  accuracy improving components of Herbie.
While ideally, Herbie would return a more accurate \textit{and} smaller
 output, Herbie's ultimate goal is to find more accurate expressions,
  even if it sacrifices AST size.
Herbie's original ruleset has been developed
  over the past 6 years by
  numerical methods experts to effectively
  accomplish this goal.
Any 
  change to these rules
  must therefore ensure that it does not make
  Herbie's result less accurate.

\autoref{fig:herbie-res} shows the results of running Herbie
  with rules synthesized by \tool.
Each box-plot corresponds to one of the four configurations.
The baseline (\lstinline{None}) and 
   \lstinline{Herbie}
  in \autoref{fig:herbie-res}'s
  accuracy and AST size plots highlight the significance of
  rational rewrites in Herbie ---
  these expert-written rules
  reduce AST size without
  reducing accuracy.
The plots for \lstinline{Ruler}
  show that running Herbie with only \tool's rational rules
   has almost the same effect on accuracy and AST size
   as Herbie's original, expert written ruleset.
The plot for \lstinline{Both} shows that running Herbie together
  with Ruler's rules further reduces AST size,
  still without affecting accuracy.
The timing plots show that adding \tool's rules to Herbie
  does not make it slower.
The baseline timing is slower than the rest because
  removing all rational
  simplification rules causes Herbie's other components
  take much longer to find the same results.\\
 
\textit{In summary, \tool's rational rewrite rules can be easily integrated 
  into Herbie, and they perform as
  well as expert-written rules without incurring any additional
  overhead.}

\paragraph{Derivability} Herbie's original rational ruleset consisted of
  52 rational rules.
\tool synthesized 76 uni-directional rational rules (50 bidirectional rules).
We compared the two rulesets for proving power, by deriving
  each with the other using the approach described in \autoref{subsubsec:derive}.
We found that Herbie's ruleset was able to
  derive 42 out of the 50 \tool rules.
It failed to derive the remaining 8.
\tool on the other hand, was able to derive all 52 rules from Herbie.
We highlight two of the 8 \tool rules that Herbie's ruleset failed to derive
that concern multiplications interaction with absolute value:
$(|a \times b| \rewritesboth |a| \times |b|)$, and
$(|a \times a| \rewritesboth a \times a)$.

\paragraph{Fixing a Herbie Bug} The above two rules found by \tool
 helped the Herbie team address a GitHub issue~\cite{herbie-bug}.
In many cases, Herbie may generate large,
 complex outputs without improving accuracy, which makes
 the program unreadable and hard to debug.
This is often due to lack of appropriate rules for
  expression simplification.
The issue raised by a user (\cite{herbie-bug}) was in fact due to the
 missing rule $(|x| \times |x| \rewritesboth x \times x)$.
The two rules above, can together, accomplish the effect of this
 rule, thereby solving the issue.
We submitted these two rules to the Herbie developers and they
 added them to their ruleset.

%% file: 06-ablation.tex
\section{Sensitivity Studies: Analyzing \ruler Framework Parameters}
\label{sec:ablation}

\begin{figure}
    \centering 
    \includegraphics[width=1.10\linewidth]{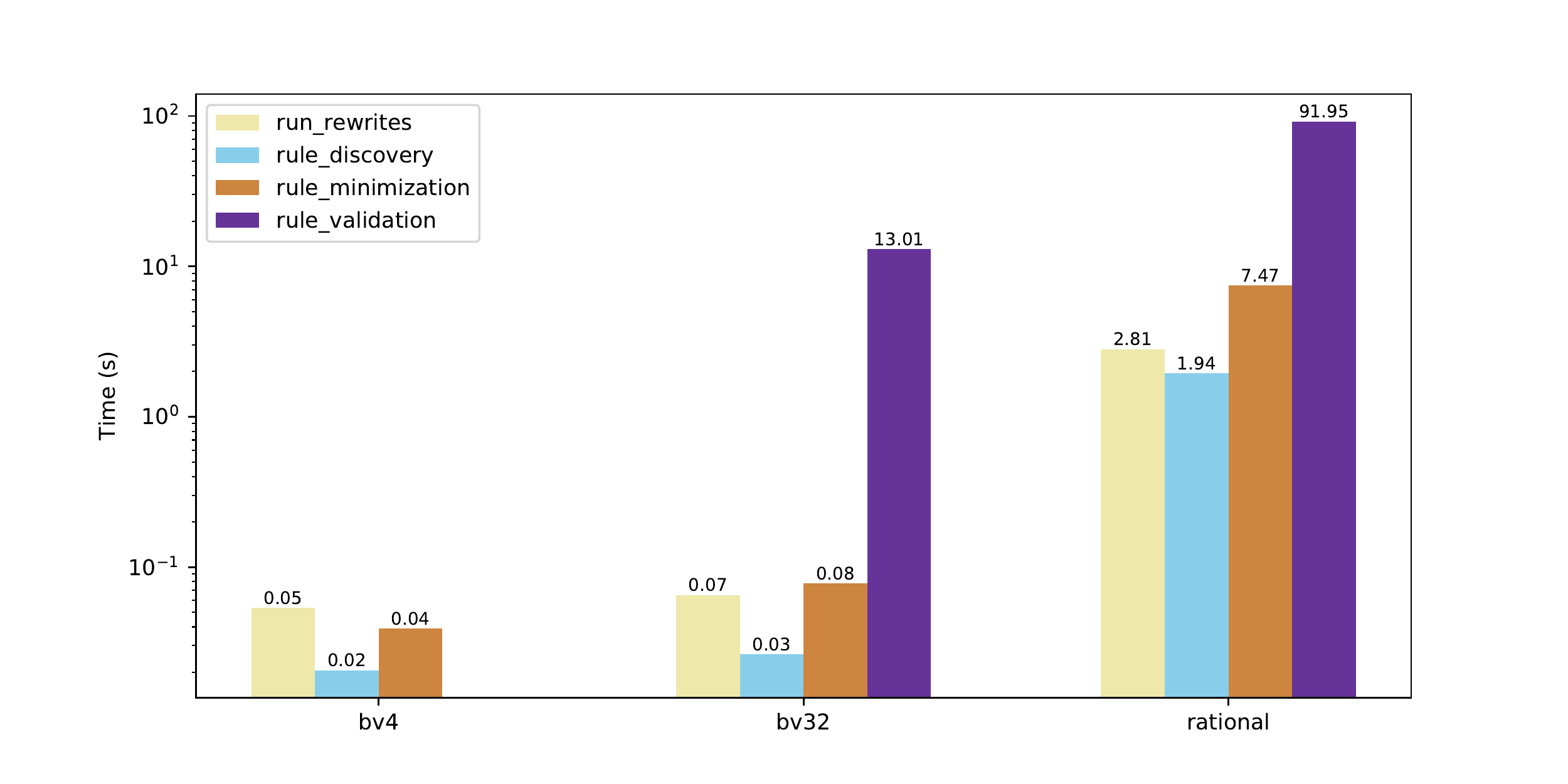}
    \caption{Search Profile.
    Time spent (in log scale)
    by \ruler in the various phases of its algorithm for
    \bfour, \bthreetwo, and \rationals.
    Most of the time is
    spent in rule validation (when applicable),
    then minimization. Notice that \bfour does not have
    an explicit validation time because the rules are correct by construction --- the \cvecs are complete.}
    \label{fig:ablation-phase-times}
\end{figure}

\begin{figure}
    \begin{subfigure}[t]{0.46\linewidth}
        \includegraphics[width=\linewidth]{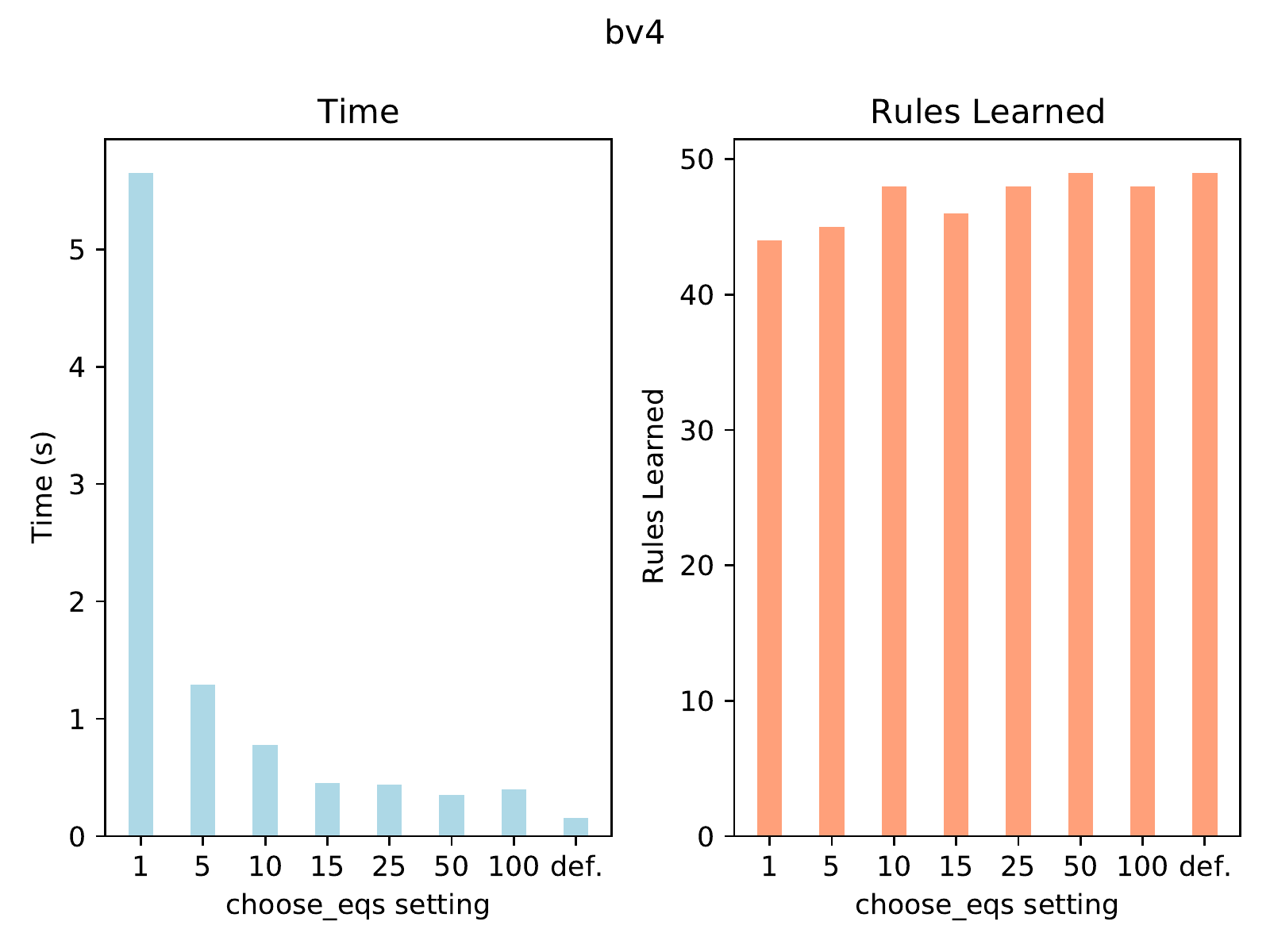}
        \includegraphics[width=\linewidth]{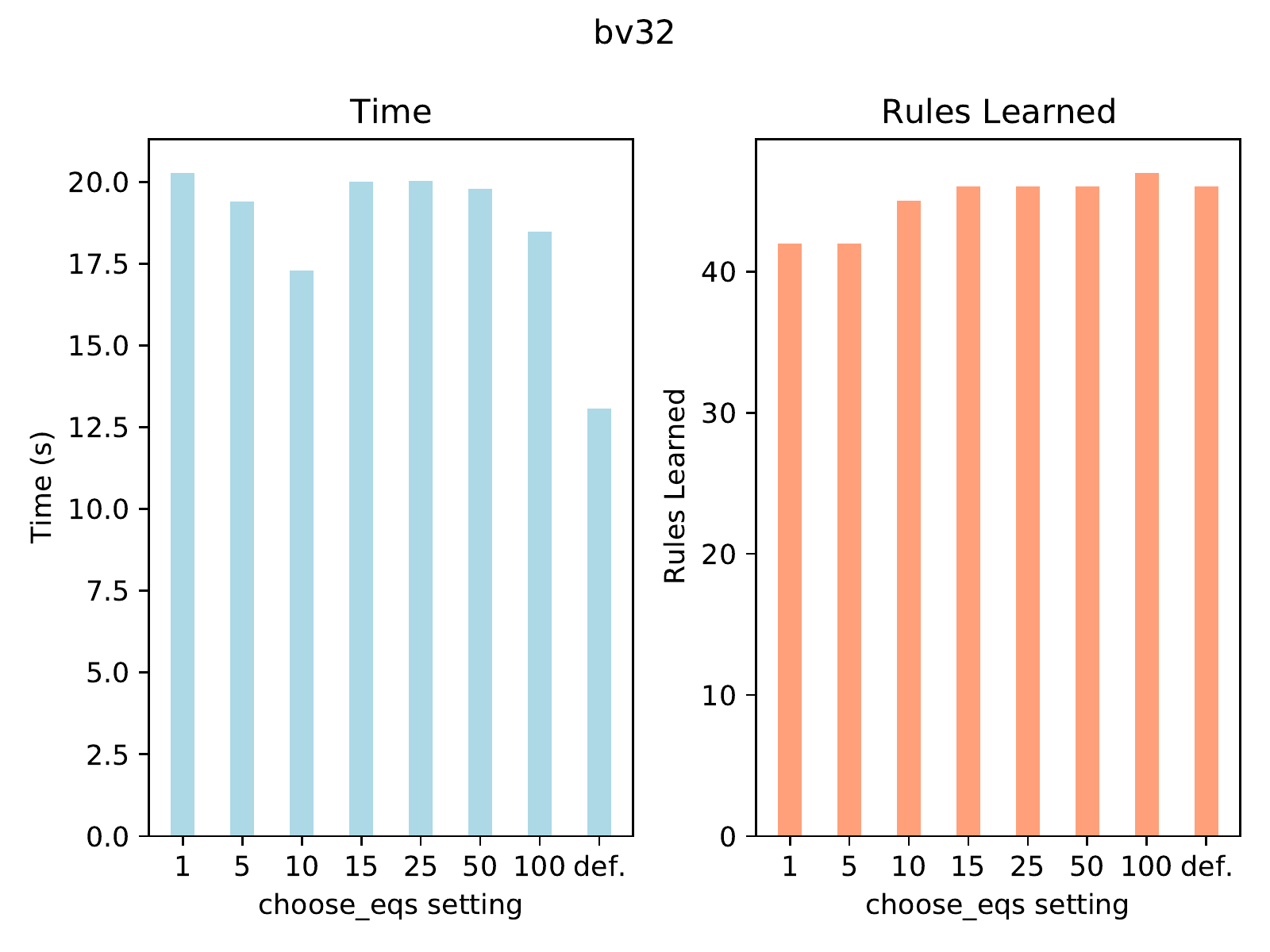}
        \includegraphics[width=\linewidth]{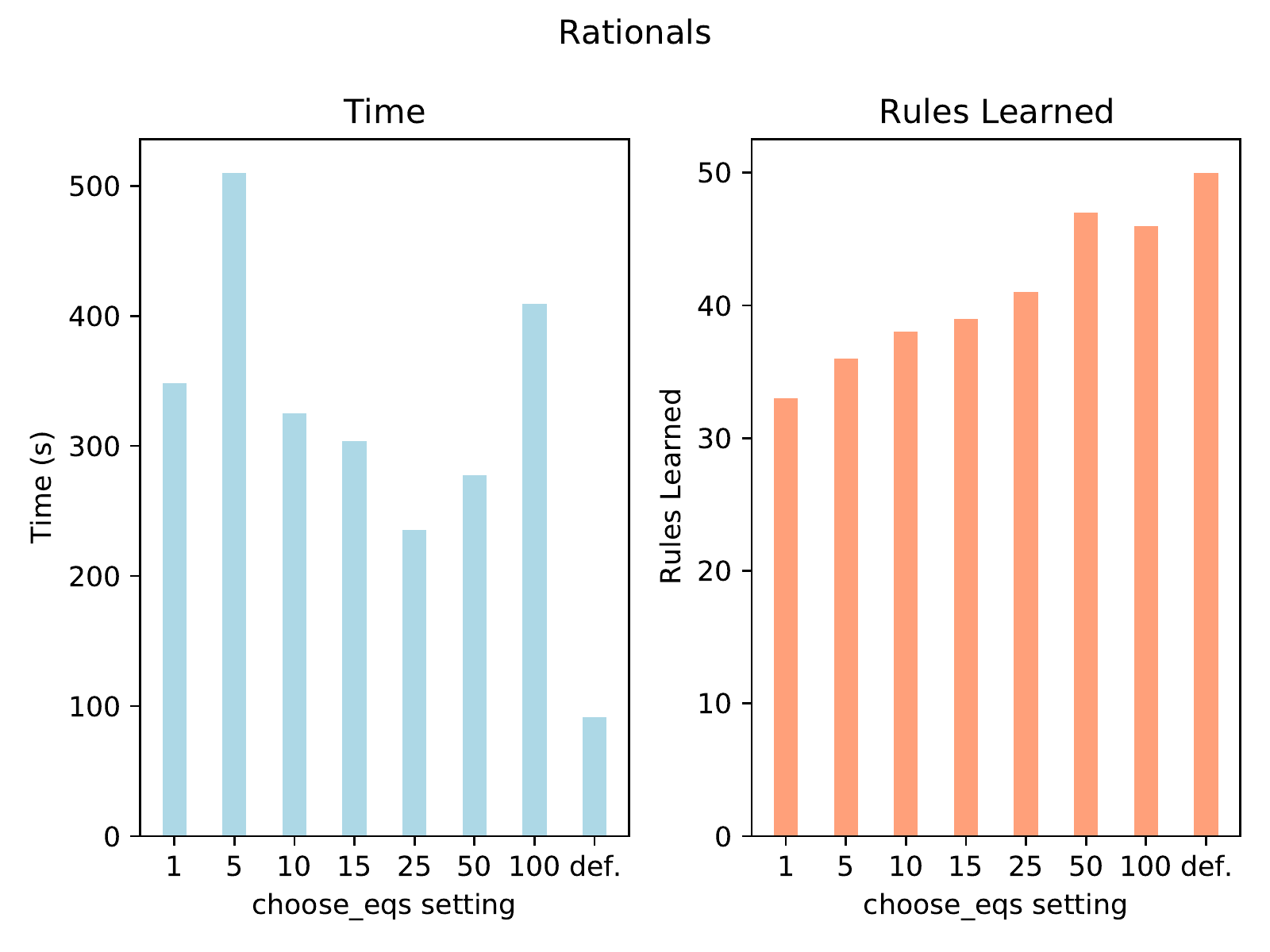}
        \caption{
          Comparison of \lstinline{choose_eqs} across values of $n$.
          \texttt{def} corresponds to $n = \infty$, \tool's default configuration.
          Larger $n$ values are generally faster but produce slightly more rules.
        }
        \label{fig:choose-eqs-rr-a}
    \end{subfigure}
    \hfill
    \begin{subfigure}[t]{0.46\linewidth}
        \includegraphics[width=\linewidth]{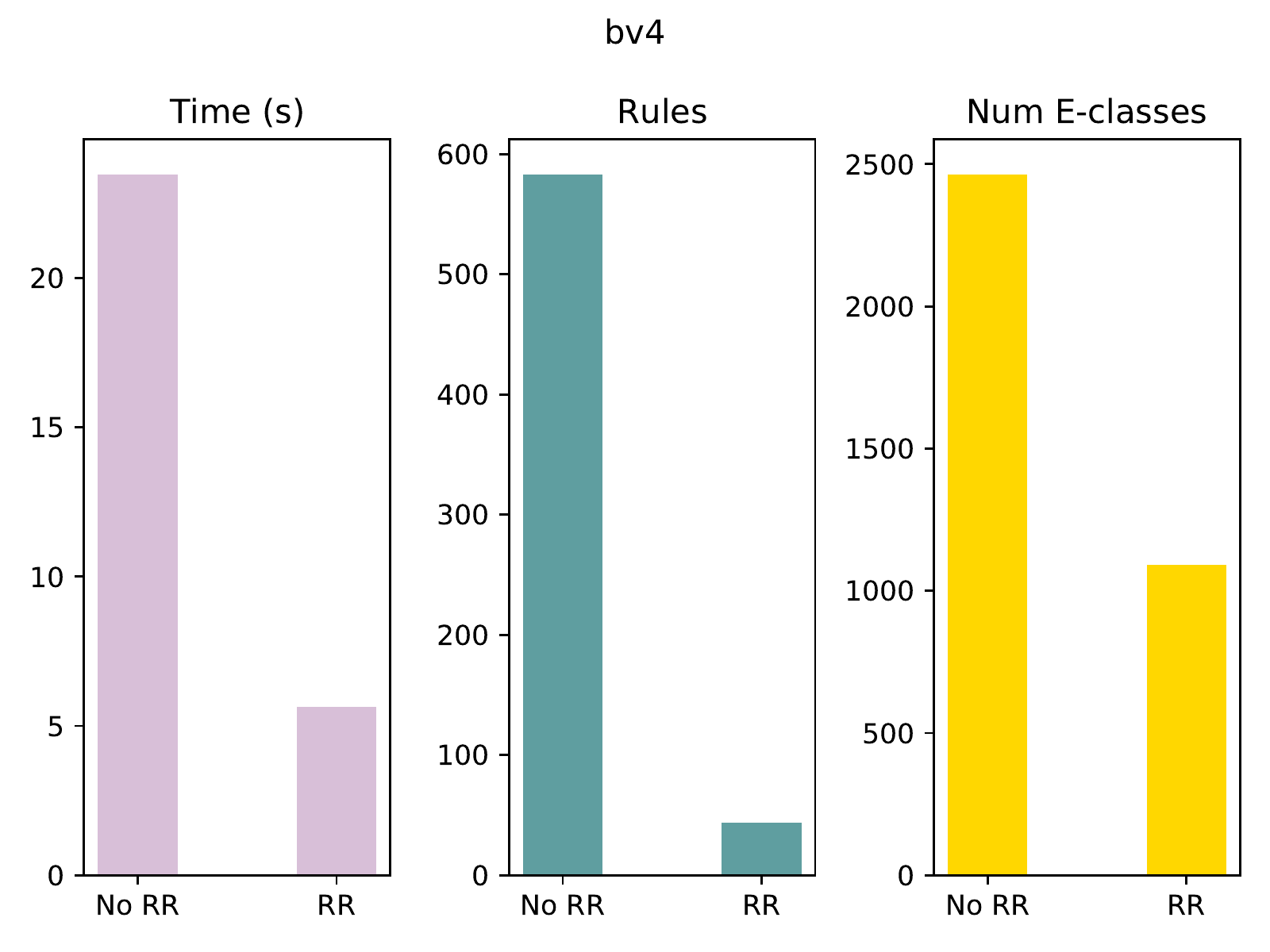}
        \includegraphics[width=\linewidth]{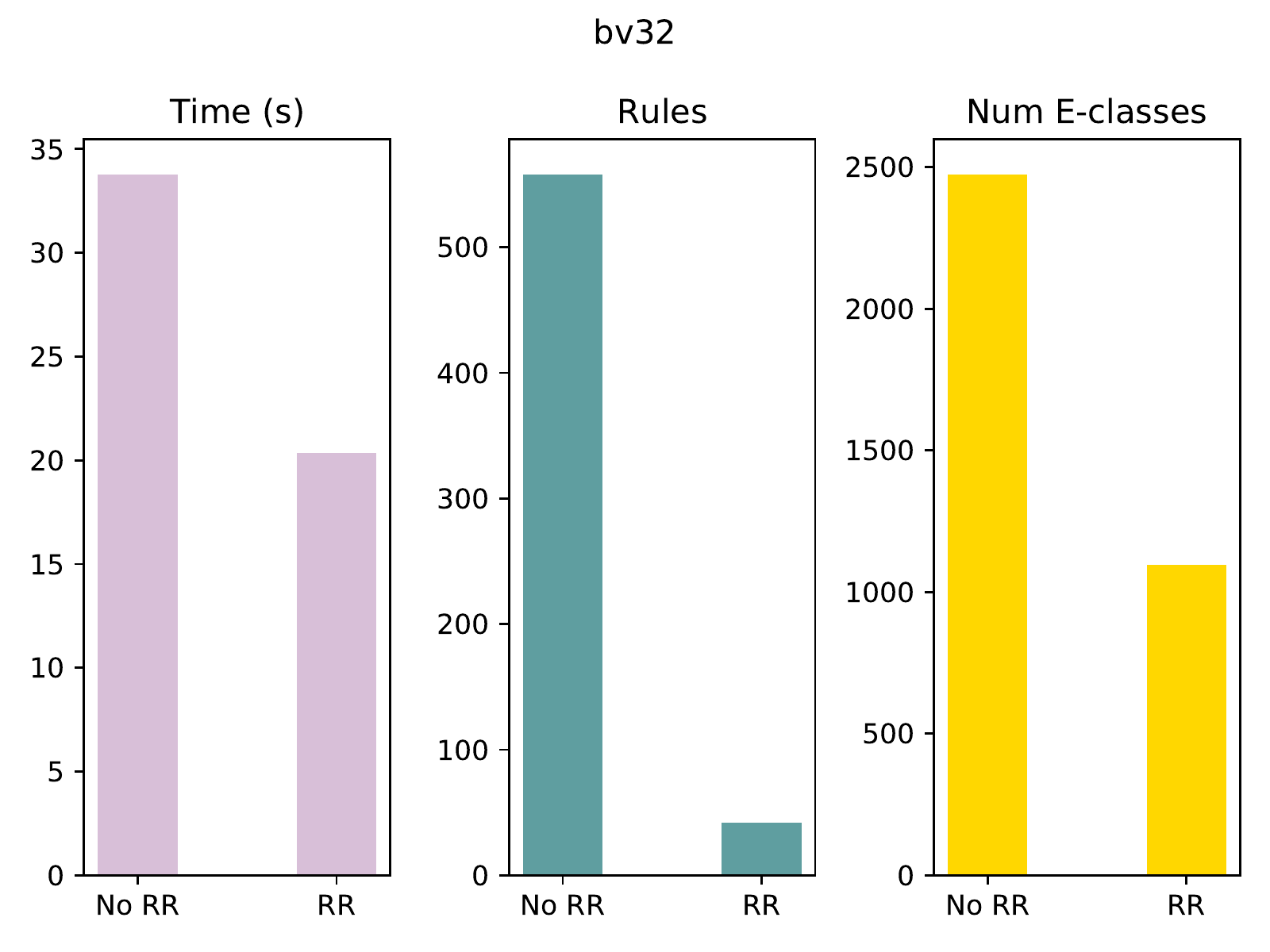}
        \includegraphics[width=\linewidth]{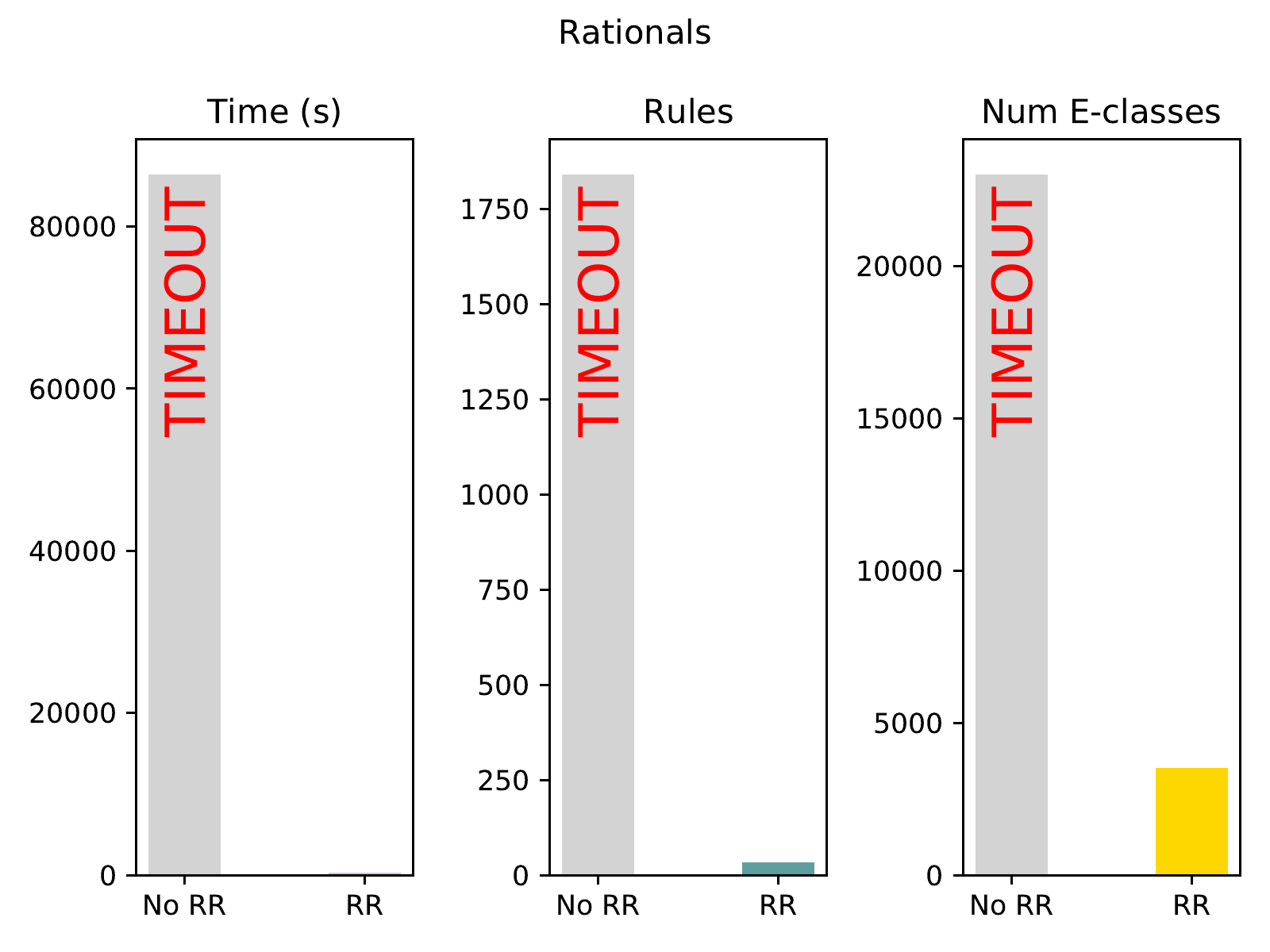}
        \caption{
          For \lstinline{choose_eqs} with $n = 1$,
          \lstinline{run_rewrites} is essential for both
          speed and synthesizing small rulesets.
          ``No RR'' is \ruler without
          \lstinline{run_rewrites}.
          For rationals, 
          with ``No RR'' did not complete in 24 hours;
          with RR, it completed in 350 seconds.}
        \label{fig:choose-eqs-rr-b}
    \end{subfigure}
    \caption{
      Comparison of \ruler's performance across variations of the search algorithm.}
  \label{fig:choose-eqs-rr}
\end{figure}

  

As a rewrite rule inference \textit{framework},
  \ruler provides parameters that
  can be varied to support different
  user-specified domains.
In particular, the
  ``learning rate'' parameter $n$
  for \mbox{\lstinline{choose_eqs}} and the choice
  of validation method present
  potential tradeoffs in terms of
  synthesis time, ruleset quality, and soundness.

In this section we evaluate how
  varying these parameters affects three
  representative domain categories:
  (1) small domains like \bfour
      where exhaustive model checking
      by complete \cvecs is feasible,
  (2) large domains like \bthreetwo
      with non-uniform behavior that
      typically require constraint solving
      for validation, and
  (3) infinite domains like \rationals
      with uniform behavior where
      fuzzing may be sufficient for validation.
We additionally profile \ruler's search
  to see the impact of \lstinline{run_rewrites}
  and compare \cvec generation strategies.


\subsection{
  Profiling \ruler Search,
  Varying \texttt{choose_eqs}, and
  Ablating \texttt{run_rewrites}}
\label{subsec:ablation}

To help guide our study of \ruler's search,
  we first profiled how much time \ruler
  spent in each phase
  across our representative domains.
\autoref{fig:ablation-phase-times}
  plots the results using
  model checking for \bfour and
  SMT for both \bthreetwo and \rationals. 
\lstinline{run_rewrites}
  (\autoref{fig:ruler-core})
  shows time compacting the
  term space with learned rules,
\lstinline{rule_discovery}
  (\lstinline{cvec_match}, \autoref{fig:ruler-core})
  shows time discovering candidate rules,
\lstinline{rule_minimization}
  (\lstinline{choose_eqs}, \autoref{fig:choose-eqs})
  shows time selecting and minimizing rules, and
\lstinline{rule_validation}
  (\lstinline{is_valid}, \autoref{fig:choose-eqs})
  shows time validating rules.
We ran each experiment in this
  section 10 times,
  for two iterations of \tool,
  and plot mean values.\footnote{
    For \rationals, \ruler iterations correspond
      to expression depth, while
      for \bfour and \bthreetwo it corresponds to
      number of connectives.
    Relative standard deviation across all
      experiments was always below 0.033.}

%

After validation (for applicable domains),
\ruler spends most of its
  search time minimizing candidate rules.
This is expected because \lstinline{choose_eqs}
  minimizes the set of candidates $C$
  by invoking equality saturation and
  $|C|$ can reach roughly $10^6$.
\ruler uses a ``learning rate'' parameter $n$
  in \lstinline{choose_eqs} to control
  how aggressively it tries to minimize rules
  (line~\ref{line:stop-n}, \autoref{fig:choose-eqs}).
When $n = 1$,
  \ruler selects only a single ``best'' rule,
  requiring more rounds of
  selecting and shrinking candidate rules.
By default, $n = \infty$,
  which causes \ruler to select a minimized
  version of that iteration's entire candidate ruleset.

\autoref{fig:choose-eqs-rr-a} shows how varying $n$
  affects overall search time and
  the resulting ruleset size:
  more aggressive minimization
  at $n = 1$ is slower but produces smaller rulesets
  relative to the default $n = \infty$.
This is expected:
  since the set of candidate rules $C$ is
  typically small compared to
  the entire term \egraph $T$,
  it is more efficient to iteratively shrink $C$
  than to repeatedly shrink $T$
  with only a few additional rules each time.
These rulesets all generally had equivalent
  inter-derivability (\autoref{sec:eval}),
  with the minimum ratio of $0.92$
  due to heuristics.

To understand how much shrinking the term \egraph $T$
  impacts search, we also compared running \ruler
  with and without \lstinline{run_rewrites}.
We set $n = 1$ for minimization as that setting
  relies the most on \lstinline{run_rewrites}.\footnote{
    We also conducted this experiment with the
    default $n = \infty$ and found
    less pronounced results as expected.}
As \autoref{fig:choose-eqs-rr-b} shows,
  \lstinline{run_rewrites} significantly improves
  search time and ruleset size while simultaneously
  requiring less space to store $T$.

\begin{table} \small
  \begin{tabular} {|lr|lr|lr|lr|lr|lr|}
    \multicolumn{2}{c}{cvec}
    & \multicolumn{2}{c}{random: 0}
    & \multicolumn{2}{c}{random: 10}
    & \multicolumn{2}{c}{random: 100}
    & \multicolumn{2}{c}{random: 1000}
    & \multicolumn{2}{c}{SMT}
    \\
    \multicolumn{12}{l}{} \\
    \multicolumn{12}{l}{4-bit Bitvector} \\
    \hline
    $C$ & 343  & --- &  --- & 49/2 & 0.1s & 49/- & 0.1s & 49/- & 0.1s & 49/- & 1.1s \\
    $C$ & 1331 & 49/- & 0.1s & 49/- & 0.1s & 49/- & 0.1s & 49/- & 0.1s & 49/- & 1.0s \\
    $C$ & 4096 & 49/- & 0.2s & 49/- & 0.2s & 49/- & 0.2s & 49/- & 0.2s & 49/- & 1.1s \\
    $R$ & 343  & 49/- & 0.1s & 49/- & 0.1s & 49/- & 0.1s & 49/- & 0.1s & 49/- & 1.0s \\
    $R$ & 1331 & 49/- & 0.1s & 49/- & 0.1s & 49/- & 0.1s & 49/- & 0.1s & 49/- & 1.0s \\
    $R$ & 4096 & 49/- & 0.2s & 49/- & 0.2s & 49/- & 0.2s & 49/- & 0.2s & 49/- & 1.1s \\
    \hline
    \multicolumn{12}{l}{} \\
    \multicolumn{12}{l}{32-bit Bitvector} \\
    \hline
    $C$ & 343  & --- & ---     & --- &    ---     & --- &  --- & --- &    ---        & 46/- & 13.6s \\
    $C$ & 1331 & 46/- & 0.1s & 46/- & 0.1s & 46/- & 0.1s & 46/- & 0.1s & 47/- & 13.2s \\
    $C$ & 6859 & 46/- & 0.2s & 46/- & 0.2s & 46/- & 0.2s & 46/- & 0.2s & 47/- & 13.3s \\
    $R$ & 343  & --- & ---     & --- &    ---     & --- &  --- & --- &    ---        & 37/- & 7.6s  \\
    $R$ & 1331 & --- & ---     & --- &    ---     & --- &  --- & --- &    ---        & 37/- & 7.7s  \\
    $R$ & 6859 & --- & ---     & --- &    ---     & --- &  --- & --- &    ---        & 37/- & 7.8s  \\
    \hline
    \multicolumn{12}{l}{} \\
    \multicolumn{12}{l}{Rational \hfill cell format: \#sound/\#unsound\quad time} \\
    \hline
    $C$ & 27  & ---  &  ---   & 50/- & 16.8s  & 50/- & 20.6s  & 47/- & 55.3s  & 50/- & 122.4s \\
    $C$ & 125 & 46/- & 33.3s  & 46/- & 32.7s  & 46/- & 32.8s  & 46/- & 34.3s  & 46/- & 38.2s  \\
    $C$ & 729 & 52/- & 342.3s & 52/- & 341.6s & 52/- & 357.2s & 49/- & 341.3s & 52/- & 349.5s \\
    $R$ & 27  & 50/- & 16.4s  & 50/- & 16.5s  & 46/- & 16.0s  & 48/- & 17.4s  & 50/- & 19.5s  \\
    $R$ & 125 & 49/- & 64.0s  & 49/- & 65.2s  & 48/- & 65.1s  & 48/- & 64.2s  & 49/- & 68.5s  \\
    $R$ & 729 & 50/- & 413.6s & 49/- & 311.5s & 49/- & 308.4s & 50/- & 414.0s & 49/- & 316.7s \\
    \hline
  \end{tabular}
  \vspace{0.2in}
  \caption{
  Comparing \cvec generation and validation strategies.
  First column shows \cvec generation approach and length:
  $C = $ Cartesian product of hand-picked values,
  $R = $ randomly sampled values.
  Middle columns correspond to validation by random testing over
  varying number of samples, and the last column is for SMT based
  validation.
  We check the rules
   for soundness with a separate, SMT-based post-pass
   and report the number of sound/unsound rules 
   and the synthesis time in seconds.
  A dashed cell indicates that \tool detected unsoundness and crashed.}
  \label{tab:valid-ablation}
\end{table}

\input{06.2-valid-ablate}
\input{06.3-domain-update}

%% file: 06.2-valid-ablate.tex
\subsection{Sensitivity Analysis for Validation Methods}
\label{subsec:soundness}

\autoref{fig:ablation-phase-times} shows that
  \ruler spends most of its search time in rule validation
  when using SMT.
To investigate the relative performance and soundness
  of other validation methods,
  we compared various strategies for constructing
  \cvecs and applying increasing levels of fuzzing
  during rule synthesis across our representative domains.
  
\autoref{tab:valid-ablation} shows that fuzzing can
  be used to synthesize surprisingly sound rulesets,
  with only a single configuration
  (\bfour, C = 343, random = 10)
  producing any unsound rules.
This is because \eqsat tends to ``amplify'' the
  unsoundness of invalid rules.
Similar to inadvertently proving \texttt{False}
  in an SMT solver, unsound rules in \eqsat
  quickly lead to attempted merges of
  distinct constants or \eclasses with
  incompatible \cvecs.
\ruler detects such unsound merge attempts and exits
  immediately after reporting
  an error to the user along with the rule that
  triggered the bogus merge (which may or may not be
  the rule ultimately responsible for introducing
  unsoundness in the \egraph).
These ``\eqsat soundiness'' crashes are indicated
  by ``---'' entries in \autoref{tab:valid-ablation}.
For the sole configuration that found unsound results
  without crashing,
  we reran the experiment with modestly increased
  resource limits and \ruler was able to detect
  the unsoundness without SMT.
Despite this encouraging result, we emphasize
  that fuzzing alone cannot guarantee soundness in general.
  
For small domains like \bfour with 3 variables,
  \tool can employ exhaustive \cvecs
  to quickly synthesize small, sound rulesets.
For larger domains like \bthreetwo,
  exhaustive \cvecs are infeasible:
  even for 2 variables they would require
  \cvecs of length $(2^{32})^2$.
Larger domains like \bthreetwo with
  subtly non-uniform behavior especially require
  verification or good input sampling since, e.g.,
  even if $x > 0$ it is possible to have $x * x = 0$.
    
To mitigate this challenge,
  \tool allows \cvecs to be randomly sampled ($R$),
  or populated by taking the Cartesian product ($C$)
  of sets of user-specified 
  ``interesting values''.
For example,
  the \ruler bitvector domains use
    values around 0, 1, \lstinline{MIN}, \lstinline{MAX},
  \rationals uses
    0, 1, 2, -1, -2, $\frac 1 2$, \ldots.

We found that for uniform domains like rationals,
  using small \cvecs with some random testing
  is sufficient for generating sound rules.
For rationals,
 the low variability in the number of rules
 learned across the different configurations
 is an artifact of \tool's minimization heuristics and
 \lstinline{cvec_matching} --- grouping
 \eclasses based on \cvecs (\autoref{subsec:cands})
 to determine which terms are matched to
 become potential rewrite rule candidates.
    
For \bthreetwo, we found that seeding
  \cvecs with interesting constants
  was more effective than random \cvecs ---
  due to the nonuniform nature of larger bitvectors,
 naively sampling random \cvec values
 was insufficient for uncovering
 all the edge cases during rule validation,
 but when unsound rules were added,
 \ruler crashed due to ``\eqsat soundiness'' violations.

%% file: 06.3-domain-update.tex
\subsection{Handling Domain Updates}
\label{subsec:updates}

An important potential application of automatic
  rewrite rule synthesis is in helping programmers
  explore the design space for rewrite systems
  in new domains or
  during maintenance of existing systems to handle
  updates when a domain's semantics evolves.
To simulate such a scenario,
  we took inspiration from the recent change
  in Halide's semantics\footnote{
    https://github.com/halide/Halide/pull/4439}
  to define $x/0 = 0$,
  and similarly changed the implementation of division
  for the \rationals domain to make the operator total.
  
Under the original \rationals semantics where
  division by zero is undefined,
  \ruler learns 50 rules in
  roughly 123 seconds with SMT validation and
  47 equivalent rules in
  roughly 21 seconds fuzzing 100 random values
  for validation (\autoref{tab:valid-ablation}).

Making division total for the \rationals domain
  in \ruler required changing a single line in the
  \rationals interpreter.
After this change, we used fuzzing with 100 random values
  to synthesize 47 rules in 18 seconds.
Since fuzzing is potentially unsound, we also
  extended SMT support in our modified version
  of \rationals with an additional 12 line change.
We then synthesized 47 rules in 59 seconds
  using SMT validation and checked that both
  the new fuzzing-inferred and new SMT-inferred rulesets
  could each completely derive the other
  (the rulesets were identical).
  
Comparing the rulesets between the
  original and updated division semantics
  revealed expected differences, e.g.,
  only the original rulesets contained
  $x/x \rewritesboth 1$ and
  only the updated rulesets contained
  $x/0 \rewritesboth 0$.
Checking derivability between the old and new
  rulesets identified 5 additional rules that
  were incompatible between the semantics,
  shedding additional light on the consequences
  of the change to division semantics.

%% file: 07-future.tex
\section{Limitations and Future Work}
\label{sec:limit}

Like any synthesis tool,
 \ruler uses limits, caps, and heuristics
 to achieve practical performance.
Of particular note are the heuristics in \lstinline{choose_eqs} ---
 \lstinline{select}
 for scoring candidate rules,
  and the \textit{step} size
 used to determine the number of
 rules to process at a time.
\lstinline{select}
  uses syntactic, size-based heuristics 
  to approximate richer concepts like subsumption.
As \autoref{sec:ablation} showed, these
  heuristics can affect
  the size of the ruleset, 
  though not significantly. 

While \ruler's \eqsat approach eliminates 
 $\alpha$-equivalent rules,
 it does not eliminate 
 $\alpha$-equivalent terms from the enumeration set $T$.
Doing so could significantly reduce the enumeration space
 and increase performance.
 
\ruler admits other implementations of term enumeration 
 via \lstinline{add_terms} (\autoref{sec:ruler}),
 but our default prototype implementation
 only explores complete enumeration.
Stochastic enumeration, potentially based on characteristics
 of a workload, could further improve scalability.
\ruler could also seed
  the initial term \egraph used for enumeration
  with expressions drawn from interesting workloads,
  e.g., benchmark suites or traces from users.
This seeding could both speed up rule inference
  and improve the effectiveness of generated rules.
Enumeration could also be limited to a semantically meaningful 
 language subset;
 for example, it is possible to learn a subset of 
 rules over reals by learning rules over rationals ---
 the latter should be faster (as rationals have fewer operators), 
 and the rational rules learned should remain sound 
 when lifted to operating over reals.

\ruler already provides limited support for
  partial operators like \lstinline{div}
  by allowing the interpreter to return a null value (\autoref{sec:ruler}).
\ruler does not, however, 
  infer the \textit{conditions}
  that ensure that partial operators succeed.
Furthermore, 
 some rewrites are total but still depend on some condition being met:
 for example  $|x|\rewritesboth x$ only when $x$ is non-negative.
An extension to \ruler could possibly to infer
  such side conditions based on
  ``near \cvec matches'' where only a few
  entries differ between two \eclasses's \cvecs,
  building on prior work~\cite{alive-infer}
  that infers preconditions for peephole optimizations.

%% file: 09-related.tex

\section{Related Work}
\label{sec:related}

This section
 discusses prior work on rewrite rule synthesis.
Most of the work in this area focuses on domain-specific
  rewrite synthesis tools, unlike \tool, which is a
  domain-general framework for synthesizing rules, given
  a grammar and interpreter.
We also focus on other framework-based approaches for rule synthesis and compare
  \ruler against them.




\subsection{Rule synthesis for SMT solvers}

Pre-processing for SMT solvers and related tools often involves
term rewriting.
Past work has attempted to automatically generate rules for such rewrites.
Recent work from \citet{sat19} is the most relevant to \ruler.
They present a partially-automated
  approach for enumerating rewrite rules for SMT solvers.
We provide a comparison with \ruler in \autoref{sec:eval}.
The main commonality between our work and theirs is
  the use of sampling to detect new equivalences;
  this is similar to \cvec matching in \ruler.
\ruler's approach is unique in its use of
  equality saturation to shrink
  both the candidate rules and the set of enumerated terms.
\citet{sat19}
  apply filtering strategies like subsumption,
  canonical variable ordering, and semantic equivalence;
  their term enumeration is based on Syntax-Guided Synthesis (SyGuS).
Their tool can be configured to use
  an initial set of rules from cvc4~\cite{cvc4} to
  help guide their search for new rules.
\ruler currently synthesizes generalized rules from scratch.
\change{Like \citet{sat19},
  \ruler generates rules that do not
  guarantee a reduction order,
  since it synthesizes rules like commutativity and associativity.
To mitigate exponential blowups when using such rules,
one approach is limiting
  the application of these rules~\cite{egg-guide}.
Prior work~\cite{szalinski} has also demonstrated
  the use of inverse transformations
  to mitigate the AC-matching problem.
\citet{julie-halide} have recently explored how to 
  infer termination orders for
  non-terminating rules,
  which could be interesting future work for Ruler as well.}

SWAPPER~\cite{swapper} is a tool for automatically generating formula
  simplifiers using machine learning and constraint-based synthesis.
SWAPPER finds candidate patterns for rules by applying machine learning
  on a corpus of formulae.
Conditions in SWAPPER are inferred by first enumerating all possible
  expressions from a predicate language. Then, the right hand side
  of the rule is synthesized by fixing the
  predicate and using Sketch~\cite{sketch}.
\citet{romano} infer reduction rules to
  simplify expressions before passing
  them to a solver,
  thereby reducing the number of queries sent to, and subsequently time spent, in the solver.
The rules are generated by symbolic program evaluation, and
  validated using a theorem prover.
\citet{nadel} used a combination of constant propagation and equivalence
propagation to speed up bit-vector rewriting in various solvers.
Several other papers~\cite{niemetz, hansen} propose algorithms and tools for
  automatically generating rules for bit-vectors.
\citet{julie-halide} recently used program synthesis
and formal verification to improve the rules in Halide.
They focus only on integer rules and
  rely on mining specific workloads to
  identify candidates for rewrites.
Preliminary experiments in \ruler indicate that supporting integers
  and even floats is achievable with
  random sampling or other validation approaches.


\subsection{Instruction Selection and Graph Substitutions}
Several tools have been proposed to automatically synthesize rewrite rules
  for instruction selectors.
\citet{buchwald-cgo18} propose
  a hybrid approach called ``iterative CEGIS'',
  combining enumeration with counter-example
  guided inductive synthesis (CEGIS) to speed up the synthesis of a rule
  library.
As the authors describe in the paper, their tool does not support division,
  and they also do not infer any rules over floats,
  since the SMT solvers they rely on are not suitable for these domains.
\ruler can be used to infer rules for domains not
  supported by SMT or that have different semantics
  because its core algorithm does not rely on SMT --- it uses SMT for verifying
  rules for domains that are supported, but \autoref{sec:ablation} shows that
  it is straightforward to use other validation techniques, or even change
  the semantics of the language and get a new set of rewrites.
\citet{dias-ramsey} proposed a heuristic search technique for automatic
  generation of instruction selectors given a machine description.
Their work uses algebraic laws to rewrite
  expressions to expand the space of expressions computable in a machine.

TASO~\cite{taso19} is a recent tool that automatically infers graph substitutions
  for optimizing graph-based deep learning computations.
TASO automatically generates rules and verifies them using Z3~\cite{z3}.
To generate the candidates, TASO enumerates expressions from a grammar up to
a certain depth and applies random testing to find equivalences, similar
to \tool.
To verify the rules, TASO uses a set of axioms that express operator properties
  in first order logic.
The axioms are used to prove that the generated rules for graph substitution
  are correct.
TASO uses subsumption to eliminate rules that are direct special cases
 of other rules.

\subsection{Theory Exploration}
\change{QuickSpec~\cite{quickspec} is a tool that automatically infers
  specifications for Haskell programs from tests
  in the form of algebraic equations.
Their approach is similar to \ruler in the sense that
  they too use tests to find potential equivalences between enumerated terms
  and filter out equations that are derivable from others.
  
Equations generated by QuickSpec have been used in
  an inductive theorem prover called HipSpec~\cite{hipspec} to
  prove other properties about Haskell programs and also
  integrated with Isabelle/HOL~\cite{hipster}.
TheSy~\cite{cav21}
  uses a symbolic equivalence technique
  for theory exploration
  to generate valid axioms for algebraic data types (ADTs).
TheSy also uses \egraphs (specifically the \egg library)
  to find equivalences and filter out redundant axioms via term rewriting.
Compared to other tools~\cite{hipster},
  TheSy typically found fewer, more powerful axioms.
Using \ruler for theory exploration~\cite{isacosy},
  especially for ADTs would be an interesting
  experiment in the future.
}

\subsection{Peephole Optimizations}
The Denali~\cite{denali} superoptimizer first showed how to use \egraphs for
optimizing programs by applying rewrite rules.
\citet{eqsat} first introduced equality saturation, generalizing
some of the ideas in Denali to optimize programs with complex constructs
like loops, and conditionals. 
Since then, multiple tools have used and further generalized
equality saturation as a technique for program synthesis, 
optimization, and verification~\cite{szalinski, spores, herbie, helm, yogo, eqsat-llvm}.
All these tools rely on the implicit assumption that the rewrite rules
will be provided to the tool.
These rulesets are typically written by a programmer and therefore
can have errors or may not be complete.
Several tools have automated peephole optimization generation~\cite{hop, bansal, alive-infer}.
\citet{bansal} presented a
  tool for automatically inferring
  peephole optimizations using superoptimization, 
  using exhaustive enumeration to generate terms
  up to a certain depth, and
  leveraging canonicalization
  to reduce the search space.
They use fingerprints to detect equivalences by
  grouping possibly equivalent terms together
  based on their evaluation
  on a few assignments.
Grouping likely equivalent terms can eliminate 
  many invalid candidate rules from even being generated in the first place.
\ruler's use of \cvecs is similar to the idea of fingerprints.

Several other papers~\cite{cond-stoke, float-stoke} have extended
and/or use STOKE for
synthesizing superoptimizations.
Alive-Infer~\cite{alive-infer} is a tool for automatically generating
pre-conditions for peephole optimizations for LLVM.
Alive-Infer works in three stages: first, it generates positive and
negative examples whose validity is checked using an SMT solver.
It then uses a predicate enumeration technique to learn predicates,
which are used as preconditions.
Finally, it uses a boolean formula learner
to generate a precondition.
\citet{alive-fp} also developed Alive-FP, a tool that
automatically verifies peephole optimizations involving floating point
computations.
Recently, \citet{alive2} published Alive2
which provides bounded, fully automatic translation validation,
while handling undefined behaviour.

%% file: 10-conclusions.tex
\section{Conclusion}
\label{sec:conclusions}

This paper presented a new technique for
  automatic rewrite rule inference using
  equality saturation.
We identified three key steps in rule
  inference and  proposed \ruler, an
  \eqsat-based framework
 that can be used to infer rule-based
 optimizations for diverse
 domains.
 
\ruler's key insight is that
  \eqsat makes each of the three
  steps of rule inference more efficient.
We implemented rule synthesis in \ruler for 
    booleans, bitvectors, and rationals.
We compared 
  \ruler against a state-of-the-art
  rule inference tool in CVC4;
  \ruler generates significantly smaller rulesets much faster.
We presented a case study showing how \ruler
  infers rules for complex domains like rationals.
Our end-to-end results show that
 \ruler-synthesized rules can replace and even surpass
 those generated by domain experts over several years.
  
We hope that this work energizes the community 
around \eqsat and incites further exciting research 
into \eqsat for rewrite rule synthesis.

